\documentclass[acmsmall]{acmart}
\renewcommand\footnotetextcopyrightpermission[1]{} 
\newcommand{\minihead}[1]{{\vspace{.45em}\noindent\textbf{#1.}}}

\newcommand{\efs}{\ensuremath{\textit{ef}_s}\xspace}
\newcommand{\efc}{\ensuremath{\textit{ef}_c}\xspace}
\newcommand{\kann}{$k-$ANN\xspace}
\newcommand{\topk}{top-\(k\)\xspace}

\makeatletter
\def\mathcolor#1#{\@mathcolor{#1}}
\def\@mathcolor#1#2#3{%
  \protect\leavevmode
  \begingroup
    \color#1{#2}#3%
  \endgroup
}
\makeatother

\definecolor{r1color}{RGB}{220,20,60}    
\definecolor{r2color}{RGB}{0,100,200}    
\definecolor{r3color}{RGB}{34,139,34}    

\usepackage{xcolor, colortbl}
\usepackage{soul}
\usepackage{marginnote}
\usepackage{ifoddpage}

\newif\ifshowmargins
\showmarginstrue  

\newcommand{\smartmargin}[2]{%
  \ifshowmargins%
      \marginpar{\raggedleft\textcolor{#1}{\textbf{#2}}}%
  \fi%
}

\newcommand{\rthree}[2]{%
  \smartmargin{magenta}{#1}%
  \textcolor{magenta}{#2}%
}

\newcommand{\eg}{\emph{e.g.,}\xspace}

\newcommand{\sys}{\mbox{\textsc{Honeybee}}\xspace}

\newcommand{\paperTitle}{\sys: Efficient Role-based Access Control for Vector Databases via Dynamic Partitioning[Technical Report]}

\newtheorem{problem}{Problem}

\newtheorem*{theorem*}{Theorem}

\newtheorem{definition}{Definition}[section]

\DeclareMathOperator*{\argmin}{arg\,min}

\usepackage{anyfontsize}
\definecolor{linkcolor}{HTML}{647382}
\definecolor{citecolor}{HTML}{647382} %
\definecolor{urlcolor}{rgb}{0.4,0.2,0.2}
\definecolor{sqlcolor}{HTML}{965d67}
\definecolor{smtcolor}{HTML}{5d968c}
\definecolor{webblue}{rgb}{0,0,.7}
\definecolor{webgreen}{rgb}{0,.5,0}
\definecolor{webbrown}{rgb}{.6,0,0}
\definecolor{notecolor}{HTML}{FFF8DC}

\usepackage{algorithm}
\usepackage{algpseudocode}
\algtext*{EndWhile}
\algtext*{EndFunction}
\algtext*{EndIf}
\algtext*{EndFor}
\algtext*{EndForAll}

\usepackage{amsmath,amsopn,amssymb}
\usepackage[cal=boondoxo]{mathalfa}
\usepackage{subcaption}
\usepackage{endnotes,microtype,xspace,graphicx,fancyvrb,multirow}
\usepackage{supertabular,booktabs}
\usepackage{array,underscore, relsize}
\usepackage{fancyhdr}
\usepackage{balance}
\usepackage{booktabs}
\usepackage{pifont}
\usepackage{listings}
\usepackage{paralist}
\usepackage{multirow}
\usepackage[scaled]{beramono}
\usepackage{tabularx}
\usepackage{multirow}
\usepackage{placeins}
\usepackage{optidef}

\usepackage{soul}
\usepackage{marginnote}
\usepackage{ifthen}

\makeatletter
\newcommand\BeraMonottfamily{%
  \def\fvm@Scale{0.85}
  \fontfamily{fvm}\selectfont
}
\makeatother

\definecolor{mymauve}{rgb}{0.58,0,0.82}

\lstdefinestyle{SQLStyle}{
  language=SQL,
  basicstyle={\scriptsize\ttfamily},
  breaklines=true,
  frame=none,
  numbers=none,
  keepspaces=true,
  captionpos=b,
  stringstyle=\color{mymauve},
  keywordstyle=\color{blue},
  commentstyle=\color{dkgreen},
}

\lstdefinestyle{ScriStyle}{
language=SQL,
basicstyle=\BeraMonottfamily\footnotesize, 
keywordstyle=\color{smtcolor}\bfseries,
morekeywords={and, or, not},
aboveskip = 0.05in,
belowskip = 0.05in,
literate = {-}{-}1, 
}

\usepackage{aliascnt}

\usepackage{semantic}
\usepackage{stmaryrd}
\usepackage{ltablex}
\usepackage{mathtools}
\usepackage{adjustbox}
\usepackage{fixltx2e}
\usepackage[group-separator={,}]{siunitx}
\usepackage{chngcntr}
\counterwithout{equation}{section}

\newcommand\blfootnote[1]{%
  \begingroup
  \renewcommand\thefootnote{}\footnote{#1}%
  \addtocounter{footnote}{-1}%
  \endgroup
}

\newcommand{\hide}[1]{}

\newcommand{\PPP}[1]{
\vspace{0.05in}
\noindent{\textit{\IfEndWith{#1}{.}{#1}{#1.}}}
}

\newcommand{\squishitemize}{
 \begin{list}{$\bullet$}
  { \setlength{\itemsep}{0pt}
     \setlength{\parsep}{0pt}
     \setlength{\topsep}{0pt}
     \setlength{\partopsep}{0pt}
     \setlength{\leftmargin}{1.95em}
     \setlength{\labelwidth}{1.5em}
     \setlength{\labelsep}{0.5em} } }

\newcounter{Lcount}
\newcommand{\squishlist}{
    \begin{list}{\arabic{Lcount}. }
   { \usecounter{Lcount}
        \setlength{\itemsep}{0pt}
        \setlength{\parsep}{3pt}
        \setlength{\topsep}{0pt}
        \setlength{\partopsep}{0pt}
        \setlength{\leftmargin}{2em}
        \setlength{\labelwidth}{1.5em}
        \setlength{\labelsep}{0.5em} } }

\newcommand{\squishend}{\end{list}}

\newcommand{\bit}{\begin{compactitem}}
\newcommand{\eit}{\end{compactitem}}
\newcommand{\ben}{\begin{compactenum}}
\newcommand{\een}{\end{compactenum}}

\newcommand{\ie}{\textit{i}.\textit{e}.,\xspace}

\definecolor{dkgreen}{rgb}{0,0.6,0}

\def\Snospace~{\S{}}

\title{\paperTitle} 

\begin{document}

\author{Hongbin Zhong}
\affiliation{%
  \institution{Georgia Institute of Technology}
  \city{Atlanta}
  \country{USA}
}
\email{hzhong81@gatech.edu}

\author{Matthew Lentz}
\affiliation{%
  \institution{Duke University}
  \city{Durham}
  \country{USA}
}
\email{mlentz@cs.duke.edu}

\author{Nina Narodytska}
\affiliation{%
  \institution{VMware Research by Broadcom}
  \city{Palo Alto}
  \country{USA}
}
\email{n.narodytska@gmail.com}

\author{Adriana Szekeres}
\affiliation{%
  \institution{Microsoft Research}
  \city{Redmond}
  \country{USA}
}
\email{aszekeres@microsoft.com}

\author{Kexin Rong}
\affiliation{%
  \institution{Georgia Institute of Technology}
  \city{Atlanta}
  \country{USA}
}
\email{krong@gatech.edu}

\begin{abstract}
Enterprise deployments of vector databases require access control policies to protect sensitive data.
These systems often implement access control through hybrid vector queries that combine nearest-neighbor search with relational predicates based on user permissions.
However, existing approaches face a fundamental trade-off: dedicated per-user indexes minimize query latency but incur high memory redundancy, while shared indexes with post-search filtering reduce memory overhead at the cost of increased latency.

This paper introduces \sys, a dynamic partitioning framework that leverages the structure of Role-Based Access Control (RBAC) policies to create a smooth trade-off between these extremes. 
RBAC policies organize users into roles and assign permissions at the role level, creating a natural ``thin waist" in the permission structure that is ideal for partitioning decisions.
Specifically, \sys produces overlapping partitions where vectors can be strategically replicated across different partitions to reduce query latency while controlling memory overhead.
To guide these decisions, \sys develops analytical models of vector search performance and recall, and formulates partitioning as a constrained optimization problem that balances memory usage, query efficiency, and recall.
Evaluations on RBAC workloads demonstrate that \sys achieves up to 13.5$\times$ lower query latency than row-level security with only a 1.24$\times$ increase in memory usage, while achieving comparable query performance to dedicated, per-role indexes with 90.4\% reduction in additional memory consumption, offering a practical middle ground for secure and efficient vector search.

\end{abstract}


\maketitle

\blfootnote{The source code is available at \url{https://github.com/rjzhb/VectorSearch-RBAC}}

\section{Introduction}
\label{sec:introduction}

Vector databases have emerged as a building block in modern applications, powering a wide variety of use cases such as in search engines, recommendation systems, and retrieval-augmented generation (RAG) pipelines driven by large language models (LLMs)~\cite{llamaindex,langchain, sawarkar2024blended, babenko2018revisiting}. 
Vector databases provide efficient vector similarity search to retrieve semantically relevant results from high-dimensional vector spaces, often implemented via approximate nearest neighbor (ANN) algorithms~\cite{malkov2018efficient, fu2019nsg, indyk1998ann, andoni2015lsh,qin2024understanding, ferhatosmanoglu2006high, gionis1999similarity}. 

As vector databases gain traction in enterprise settings, particularly in applications like retrieval-augmented generation (RAG) pipelines powered by large language models, enforcing appropriate access controls has become a critical challenge~\cite{supabase2025}. Role-based access control (RBAC) is a widely adopted framework for managing access to sensitive data~\cite{ferraiolo1992rbac,sandhu1996rbac,colombo2010rolemining,erbac,sandhu1998role}. RBAC simplifies permission management by grouping users into roles (e.g., HR, Finance, Engineering) and assigning data access permissions to these roles, rather than managing permissions at the individual user level. While RBAC is well-established for traditional relational databases, its implementation in vector databases introduces unique complexities. In relational databases, RBAC can be enforced through simple query predicates (e.g., "SELECT * FROM documents WHERE user=Alice"). However, vector databases must integrate these relational predicates with vector search on ANN indexes, requiring hybrid queries that simultaneously satisfy both vector similarity and access control constraints.

\begin{figure}[t]
    \centering
    \includegraphics[width=0.8\linewidth]{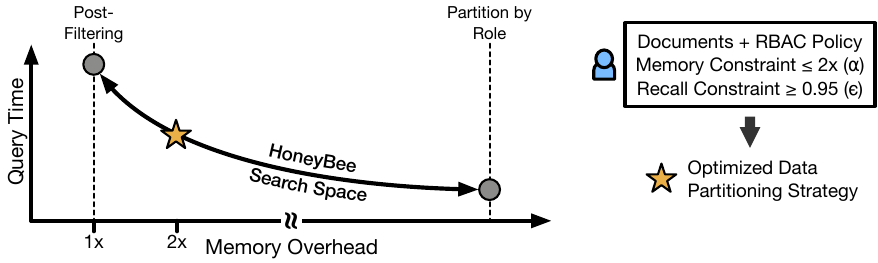}
    \caption{
   Given RBAC policies, memory and recall constraints, \sys optimizes for a data partitioning strategy that achieves a balanced trade-off between query latency and memory overhead.}
    \label{fig:fig1}
\end{figure}

Access control can be readily integrated with vector databases using one of the two approaches: 
(1) dedicated indices for each user (or role), or (2) post-filtering on a single, shared index.
These approaches represent two ends of a spectrum in the trade-off between memory overhead and query performance, as shown in \autoref{fig:fig1}.
The dedicated indices approach creates separate indices for each user or role, ensuring that queries only access authorized data. This eliminates the need for runtime filtering and enables fast query execution. However, it incurs significant memory overhead, as vectors accessible to multiple users or roles must be duplicated across partitions. For example, in our experiments, even partitioning by roles (rather than individual users) resulted in memory overheads that were 7$\times$ larger than a unified index. 
In contrast, the post-filtering approach constructs a single unified index and applies access control filters after performing the ANN search. Although this minimizes memory overhead by avoiding data duplication, it suffers from poor recall and performance when filters are highly selective (\ie users can access only a small fraction of the data). In such cases, most search results are discarded, requiring the system to expand the search scope to achieve acceptable recall, which increases query latency. Recent works have explored specialized index structures to better support vector similarity search with filtering conditions~\cite{wu2022hqann,patel2024acorn,gollapudi2023filtered}.
However, these methods still operate within the constraints of a single index and do not fully exploit the potential benefits of storing redundant vector copies across indices.

In this work, we propose \sys, a \emph{dynamic partitioning framework} that efficiently implements RBAC in vector databases. 
Our key insight is that the structure of RBAC policies can be leveraged to design a hybrid approach that balances the strengths of dedicated indices and post-filtering.
Enterprise RBAC deployments typically feature far fewer roles than users (e.g., tens to hundreds of roles versus tens of thousands of users). 
Moreover, role definitions tend to remain stable over time, even as user assignments change frequently~\cite{ferraiolo1992rbac,sandhu1996rbac,colombo2010rolemining,roleminer,sandhu1998role}. 
Empirically, we find that partitioning by user (or unique combinations of roles) significantly increases memory overhead compared to partitioning by roles, while offering diminishing returns in query latency improvements (\eg $>50\times$ increase in memory for $<2\times$ improvement in query latency).
We exploit these properties by treating roles as the finest level of partitioning granularity. 

Specifically, \sys introduces analytical models to quantify the expected search performance and recall with respect to key factors, such as partition size, average filter selectivity, and index-specific parameters that control the trade-off between search performance and recall. 
Based on these models, \sys formulates the partitioning strategy as a mixed-integer nonlinear programming (MINLP) optimization problem. 
Since solving this problem is NP-hard, \sys further introduces a greedy algorithm that generates a spectrum of partitioning strategies under different memory constraints, ranging from post-filtering to dedicated indices per role.
The partitioning strategy operates independently of the underlying index structure, making \sys applicable across different vector search indices--requiring only the corresponding performance model.
For example, specialized hybrid search indices can be used on individual partitions to further improve post-filtering performance within each partition. 
This flexibility allows \sys to create a spectrum of solutions that balance memory overhead and query performance.

In summary, our key contributions are as follows:
\squishitemize
    \item We identify the unique challenges of implementing access control in VectorDBs and analyze the limitations of existing approaches, including post-filtering and partitioning.
    \item We propose \sys, which integrates RBAC with optimized data partitioning strategies, striking a balance between memory and query efficiency.
    \item We demonstrate the adaptability of our framework to various operational scenarios, including its compatibility with hybrid search methods such as ACORN.
    \item We evaluate our approach on real-world datasets with diverse permission structures, highlighting its practical applicability compared to state-of-the-art baselines. Using pgvector with the HNSW index, \sys reduces memory redundancy compared to dedicated indices per role and achieves up to 13.5$\times$ faster query speeds than post-filtering on shared index (implemented via PostgreSQL's row-level security feature), with only 1.24$\times$ memory increase, while achieving comparable query performance to role partition with 90.4\% reduction in additional memory consumption.
\squishend

\section{Background}
In this section, we provide background on role-based access control (RBAC) (\S~\ref{sec:rbac-bg}) and hybrid vector search (\S~\ref{sec:index-bg}). 

\subsection{Role-Based Access Control}
\label{sec:rbac-bg}

The simplest way to enforce access control is via access control lists (ACLs), which specifies for each protected resource (e.g., a patient's record), a list of users that have access to that object. 
However, ACLs face several challenges. 
ACL directly associate users with permissions, which is hard to maintain when dealing with a large number of users and permissions that need constant updating.
Furthermore, organizations typically need to specify access policies based on user functions or roles within the enterprise. 
For example, the principle of least privilege, which states that users and applications should only have access to the data and operations necessary for their jobs, is burdensome to implement directly on top of ACLs.

Role-Based Access Control (RBAC) emerged as a solution to address these aforementioned challenges in enterprise settings~\cite{ferraiolo1992rbac,sandhu1996rbac,
colombo2010rolemining}. By introducing roles as an intermediary layer between users and permissions, RBAC simplifies the management of complex access policies and reduces administrative overhead.

\begin{definition} 
$\gamma = \langle U,R,D,\phi_{UA},\phi_{PA} \rangle$ defines a basic RBAC system, where
\begin{itemize}
\item $U$, $R$ and $D$ are the sets of users, roles, and documents in the systems, respectively. 
\item $\phi_{UA}: U \rightarrow 2^R$ defines the many-to-many relationship between users and roles. 
\item $\phi_{PA}: R \rightarrow 2^D$ defines the many-to-many relationship between roles and documents. 
\end{itemize}
In this system, a user $u_i$'s authorized permissions are determined by the union of permissions acquired through their assigned roles: $auth(u_i) = \bigcup_{r \in \phi_{UA}(u_i)} \phi_{PA}(r)$.
\label{def:rbac}
\end{definition}

The RBAC system can also incorporate a partial order of roles to support role hierarchies. For example, in a healthcare setting, rather than granting physicians blanket access to all patient record data, RBAC allows administrators to define a hierarchy of roles based on the physician's specialization, and restrict access to only those fields relevant to a particular type of physician's practice.

Relational databases such as SQL Server and PostgreSQL provide Row-Level Security (RLS) as a built-in security feature to control access to rows in a database table \cite{satoricyberrls,microsoftrls}. RBAC policies can be easily implemented using a set of tables that mirror the main RBAC components. When a user attempts to access a protected resource, the database system performs a series of JOIN operations across these tables to determine if the access should be granted.
For example, \autoref{lst:rls-code} checks if a user has permissions to specific rows in the \texttt{PatientRecords} table by joining the user-role and role-permission assignment tables. 


\begin{figure}[h]
\centering
\begin{tabular}{c}
\begin{lstlisting}[style=SQLStyle, morekeywords={POLICY}]
CREATE POLICY access_policy ON PatientRecords 
FOR SELECT USING (
EXISTS (
        SELECT 1
        FROM PermissionAssignment pa
            JOIN UserRoles ur ON pa.role_id = ur.role_id
        WHERE pa.record_id = PatientRecords.record_id
          AND ur.user_id = current_user::int
));
\end{lstlisting}
\end{tabular}
\caption{RBAC policy implemented via Row-Level Security. \label{lst:rls-code}}
\vspace{-1em}
\end{figure}

\subsection{Relationship with Hybrid Search}
\label{sec:index-bg}
Hybrid search refers to queries jointly searching vector data (embedded images and text) and structured data (attributes and keywords).
For example, access control policies for vector databases can be implemented as hybrid search queries, where users retrieve documents relevant to their query (measured by vector distance) among those accessible (structured filtering condition).

Existing hybrid search solutions fall into two categories.
First, index-based solutions (e.g., ACORN~\cite{patel2024acorn}, Filtered-DiskANN~\cite{gollapudi2023filtered}, Curator~\cite{jin2024curator}) modify existing vector indices to handle filtered queries more efficiently. These approaches adapt both index construction and querying processing to improve post-filtering performance \emph{within a single index structure}. For example, ACORN~\cite{patel2024acorn} enhances Hierarchical Navigable Small Worlds (HNSW)~\cite{malkov2018efficient}, state-of-the-art graph-based index for approximate nearest neighbor search, by integrating filtering conditions directly into index construction. 
By expliciting considering filtering predicate selectivity during graph construction, ACORN can improve connectivity and achieve better query recall and latency versus post-processing approaches that apply filters after vector search.

Second, \emph{partition-based solutions} (e.g., HQI~\cite{hqi}, our approach) partition datasets according to query predicates and build separate indices for each partition. Our framework falls into this category and is complementary and largely orthogonal to index-based methods--we address how to optimally distribute data across indices based on access policy structure, while index-based methods focus on enhancing individual index structures. 
In the evaluation (\S~\ref{sec:exp-index}), we show our approach complements index-based methods and integrates with existing hybrid search indexes to further enhance performance for permission-aware vector search.

We provide a more detailed comparison with prior works in both categories in the related work~section (\S~\ref{sec:related}).


\begin{figure*}[t]
    \centering
    \includegraphics[width=1.0\linewidth]{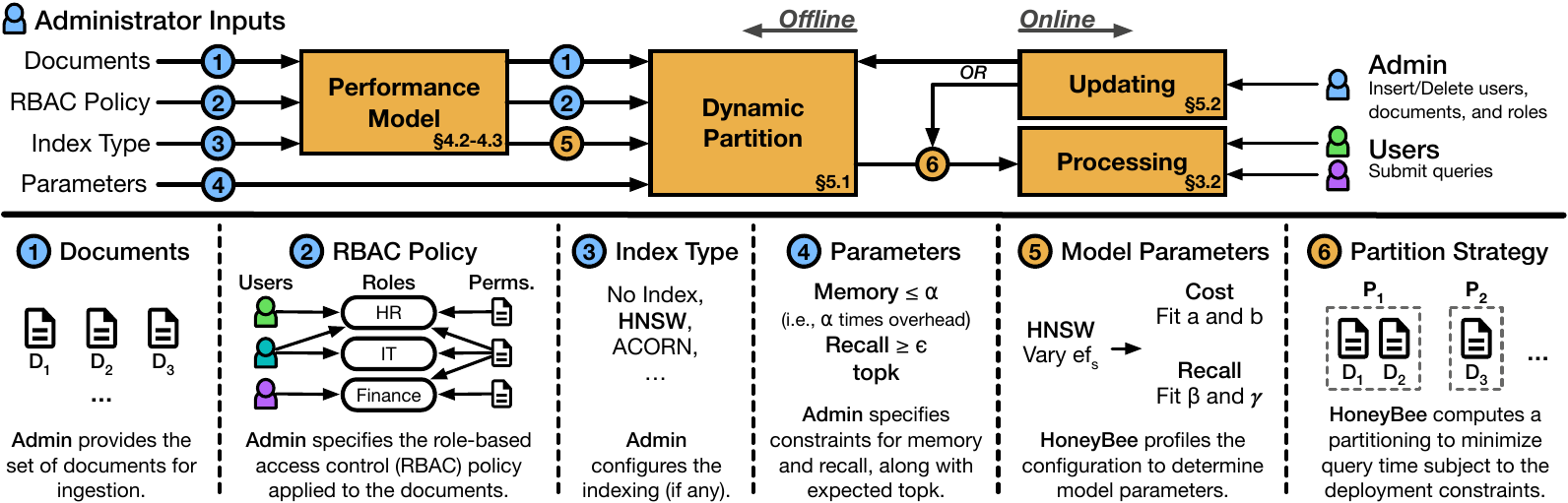}
    \caption{Overview of \sys's workflow. Yellow elements represent main components as well as inputs and outputs of \sys, while blue elements represent configurable components by administrators, such as documents, RBAC policies, index types, and constraints.}
    \label{fig:overview}
\end{figure*}

\section{Overview}
In this section, we provide an overview of \sys, including the problem statement (\S~\ref{subsec:problemstatement}) and main system components (\S~\ref{sec:sysoverview}). 

\subsection{Problem Statement}
\label{subsec:problemstatement}
Consider a vector database system managing document access with role-based permissions. Let $D = \{d_1, d_2, ...\}$ be the set of all managed documents. Each document $d_i \in D$ represents an atomic unit for permission assignment and may contain either a single vector (e.g., embeddings for a single image) or multiple vectors (e.g., embeddings for paragraphs within a webpage).

Let $U = \{u_1, u_2, ...\}$ be the set of users, and \(R = \{r_1, r_2, ...\}\) be the set of roles (e.g., "manager", "HR"). As discussed in Definition~\ref{def:rbac}, RBAC policies can be defined by two mappings, $\phi_{UA}: U \rightarrow 2^R$ (assigns users to roles), and $\phi_{PA}: R \rightarrow 2^D$ (grants roles access to documents). $auth(u_i)$ is the set of documents accessible to user $u_i$.

The system processes queries in the form of $q_i=(u_{q_i}, v_i)$, containing a user $u_{q_i} \in R$ and a query vector $v_i$. The vector database's goal is to retrieve the top-k most relevant documents based on vector similarity, subject to the users' access permissions. 

To optimize for search performance, the system can be configured using different partitioning strategies, where a separate index is built on each partition. 
A configuration is represented as a non-disjoint partitioning $\Pi = \{\pi_1, \pi_2, ...\}$, where $\cup_{i} \pi_i = D$ and $\forall i: \pi_i \subseteq D$ and $\pi_i \neq \emptyset$. 
We use the term ''partitioning'' in this context to refer to the creation of physically separate indexes that may have data overlap, rather than a strict mathematical disjoint partitioning.
To ensure high query performance, we assume that all indices are stored in memory, whether using a single shared index or multiple indices built on each partition.

The set of partitions that contain documents accessible to role $r_i$ is defined as: $P_r(r_i, \Pi) = \{j \mid \pi_j \in \Pi,  \pi_j \cap \phi_{PA}(r_i) \neq \emptyset\}$.
The set of partitions that contain documents accessible to user $u_i$ is the union of partitions accessible to each of its roles: $P_u(u_i, \Pi) = \bigcup_{r \in \phi_{UA}(u_i)} P_r(r, \Pi)$.


Configurations are evaluated on three dimensions:
\squishitemize
\item \textbf{Memory overhead}: $\frac{\sum_{i}|\pi_i|}{|D|}$, the ratio of total required memory of the configuration to the minimum memory requirement of using a single shared index. 
We use partition size as a proxy for the memory overhead, since index size is proportional to the number of vectors stored.
We focus on \emph{physical partitioning}, where each partition maintains its own copy of the vector data. This design follows the implementation of vector indexes such as HNSW, which physically co-locates vector data and index metadata (\eg graph structure) to enable low latency queries. \S~\ref{sec:eval-logical} provides a comparison with logical partitioning. 

\item \textbf{Search quality}: $R(\Pi, u_i)$, the search recall for user $u_i$ is defined as the ratio between (1) results from indexed search filtered by $u_i$'s access permissions and (2) the ground-truth \topk results from exhaustive search followed by access control filtering.
\item \textbf{Search performance}: $C(\Pi, u_i) = \sum_{j \in P_u(u_i, \Pi)} c(\pi_j)$, where $c(\pi_j)$ is the cost of retrieving the \topk nearest neighbors for the query in partition $\pi_j$. $c(\pi_j)$ depends on factors such as partition size $|\pi_j|$, index types, and index parameters.
\squishend
In the following sections, we will introduce analytical models to approximate the query performance and recall models. 

\begin{problem}
\label{def:prob}
Given a set of documents $D$, RBAC policies, memory constraint $\alpha (\geq 1)$, and recall constraint $\epsilon (<1)$, find a partitioning $\Pi=\{\pi_1, \pi_2, ...\}$ that minimize the average query latency over users:
\begin{align*}
\underset{\Pi}{\text{minimize}} & \quad \frac{1}{|U|}\sum_{i=1}^{|U|} C(\Pi, u_{i}), 
\text{s.t.} \frac{\sum_{i} |\pi_i|}{|D|} \leq \alpha,  \frac{1}{|U|}\sum_{i=1}^{|U|}R(\Pi, u_i) \geq \epsilon
\end{align*}

Alternative objectives are also available, such as the average query latency over a query workload (weighted version of the above), or the average query latency over all roles.
\end{problem}

\subsection{System Overview}
\label{sec:sysoverview}

\sys operates in two phases: an offline phase for data partitioning and indexing, and an online phase for query processing. \autoref{fig:overview} provides an overview of the system workflow. 

\minihead{Offline} 
In the offline phase, \sys solves the constraint optimization problem (Problem 1) to determine the optimal partitioning strategy.
Depending on the memory constraints, this strategy can range from a single shared partition (equivalent to post-filtering on a shared index) to role-specific partitions (dedicated indices approach), or more commonly, utilize a hybrid approach with overlapping partitions based on access patterns. 

For each partition $\pi_i \in \Pi$, \sys builds a separate similarity search index. The index type is configurable by users, with options including no index (exhaustive search), vector indices for \kann search (\eg HNSW, IVF-PQ), or specialized indices for hybrid search (\eg ACORN). 
Along with partitioning, \sys determines the index-specific parameter that controls search depth (\ie number of candidate results considered during search) and affects the search latency and recall during query time.
The search depth is influenced by the partitioning design. 
When all documents are in one partition, high search depth is required to ensure sufficient results remain after permission filtering. However, when documents are distributed across multiple partitions, each partition contains a higher density of accessible documents for its intended users, allowing \sys to use a lower search depth while achieving the target recall. Our analytical model for query recall (Eq~\ref{eq:recalldef}) provides an initial recommendation for these parameters, which users can further finetune to meet specific requirements if needed.  

Since partitions may overlap, the set of partitions containing documents accessible to a user ($P_u(u_i, \Pi)$ defined in \S~\ref{subsec:problemstatement}) could include redundant partitions. 
To optimize query processing, \sys pre-computes and maintains a routing table, $P^{*}(u_i, \Pi)$, from each unique combination of roles (or each distinct user) to their minimal required set of partitions: 
\begin{equation}
P^{*}(u_i, \Pi) = \argmin_{S \subseteq P_u(u_i, \Pi)} \sum_{j \in S} c(\pi_j), \text{ s.t. } \bigcup_{j \in S} \pi_j \supseteq auth(u_i)
\label{eq:trackermin}
\end{equation}
where $auth(u_i)$ is the set of documents accessible to user $u_i$ (Def~\ref{def:rbac}). For role-level analysis, $P^{*}(r_i, \Pi)$ is defined analogously by substituting $auth(u_i)$ with $auth(r_i)$ in Eq.~\ref{eq:trackermin}.
Once precomputed, locating the required partitions during query time requires a single key-value lookup ($O(1)$ operation). The routing table itself is extremely lightweight— even with ten thousand entries (each storing a few integer partition IDs), it only occupies a few hundred kilobytes, negligible compared to the gigabyte-scale vector indexes. 

To identify the covering set, we iteratively choose the partition that covers the largest number of uncovered documents; when multiple partitions provide equal coverage, we select the smaller partition to reduce scan cost.

\minihead{Online} 
When processing a query $q_i = (u_{q_i}, v_i)$, \sys first identifies the relevant partitions using the precomputed mapping $P^{*}(u_{q_i}, \Pi)$. For each identified partition, the system performs vector similarity search using the query vector $v_i$ with a vector index. 
The search process is controlled by index-specific parameters that determine how many candidates to examine (e.g., \emph{ef\_search} for HNSW, \emph{nprobe} for IVF-PQ).
After vector search, access control is applied to filter out unauthorized documents through post-filtering (\autoref{lst:rls-code}). 
Finally, \sys merges the filtered results from all searched partitions, sorts them by similarity score, and returns the global top-$k$ results.

\section{Analytical Modeling}
\label{sec:model1}
In this section, we introduce the analytic models for the vector search performance and recall, using HNSW index with post-filtering as a concrete instantiation of the framework.

\subsection{Background: HNSW Index}
HNSW is among the most widely adopted graph-based indexes for ANN search~\cite{malkov2018efficient}.
It organizes vectors as nodes in proximity graphs where edges connect similar vectors.
During search, the algorithm starts from predefined entry points and greedily moves to neighboring nodes closer to query vectors.
HNSW performance is governed by three key parameters: \(M\),  $ef\_construction$ ($\efc$), and $ef\_search$ ($\efs$). 
$M$ and $ef_c$ are construction-time parameters defining the structure and quality of the HNSW graph.
\(M\) controls link numbers each node maintains, affecting memory usage and search efficiency.
$\efc$ determines neighbors considered when inserting new points, influencing indexing speed and accuracy. 
In contrast, $ef_s$ is a query-time parameter that can be adjusted without rebuilding the index, making it the primary lever for latency–recall trade-offs. 
$\efs$ controls search depth by determining priority queue size used to manage candidate nodes during queries. 
Higher $\efs$ values improve recall at increased query latency.

In this study, we use the recommended default settings for $M$ and $ef_c$ following common practices~\cite{hnswdefaults,hnswparam}. Our analytical model treats $M$ and $ef_c$ as system-level constants while focusing on jointly optimizing partitioning and the query-time parameter $ef_s$.
Tuning $M$ and $ef_c$ controls the graph’s connectivity and construction-time search quality, which affects both index size (and thus memory constraints) and recall (and thus performance models).
Incorporating their tuning into \sys’s offline optimization could expand the design space and enable partitioning strategies with improved memory–latency tradeoffs.
For tractability, we also use a single global $\efs$ across all partitions. While this simplifies the model, it limits the role of $\efs$ in the recall–latency tradeoff, as different partitions may benefit from different query-time settings depending on their size and selectivity. 
Allowing per-partition $\efs$ could substantially enlarge the online tuning space and lead to latency gains in highly heterogeneous partitions.
We leave such multi-parameter and per-partition tuning to future work.

The following sections introduce analytical models characterizing the relationship between \efs, query recall, and latency.


\subsection{Model for Search Performance}
\label{sec:model2}

The query time in HNSW is influenced by two main factors: the traversal of the hierarchical graph and the cost of vector similarity calculations. The traversal complexity is approximately \(O(\log n)\), where \(n\) is the number of points in the graph~\cite{malkov2018efficient}. 
The parameter \efs directly impacts query time by determining the number of candidate nodes visited during the search. 

We model the query time for a partition as $c(\pi_i, \efs) = \log(|\pi_i|) \cdot f(\text{ef}_s)$, where $f(\text{ef}_s)$ represents the relationship between search queue size and query time. 
Empirical analysis demonstrates that $f(\text{ef}_s)$ exhibits a linear relationship
with $\text{ef}_s$ which we model as $f(\text{ef}_s) = a \cdot \text{ef}_s + b$. Here, $a$ and $b$ are system-dependent parameters influenced by hardware capabilities, software optimizations, and dataset characteristics such as intrinsic dimensionality.

Overall, the query cost is a function of the partitioning scheme $\Pi$, the index specific parameter $\efs$, and the query itself. 
We consider two types of query costs: user-level ($C_u$) and role-level ($C_r$). 
For a given user $u_i$, the total query cost $C_u(\Pi, u_i, \efs)$ must account for all partitions that contain documents accessible to that user, as defined by $P^{*}(u_i, \Pi)$ in Eq~\ref{eq:trackermin}:

 \begin{equation}
C_u(\Pi, u_i, \efs) = \sum_{j \in P^{*}(u_i, \Pi)} \log(|\pi_j|) \cdot (a \cdot \text{ef}_s + b) 
\label{eq:usercost}
 \end{equation}
Similarly, the role-level query cost is defined as 
\vspace{-0.5em}
\begin{equation}
C_r(\Pi, r_i, \efs) = \sum_{j \in P^{*}(r_i, \Pi)} \log(|\pi_j|) \cdot (a \cdot \text{ef}_s + b) 
\label{eq:rolecost}
 \end{equation}
Note that since different roles could be mapped to the same partitions via $P^{*}(r_i, \Pi)$, the user-level cost is not simply a weighted average of the role-level cost.

To fit \(a\) and \(b\), we use an RBAC permission generator (\S~\ref{sec:generators}) to create a workload where each user maps to one role, and one partition is created per role.
Although users have a 1:1 mapping with roles, different role-based partitions may still overlap.
Here, \(\text{P}^{*}(r_i, \Pi) = \pi_{r_i}\), and \(C_u(\Pi, u_i, \efs) = C_r(\Pi, r_i, \efs) = \log(|\pi_{r_i}|) \cdot (a \cdot \efs + b)\). 
We generate 1000 queries, test multiple \efs values, and derive an average query time per \efs. This allows us to compute \(\frac{\text{querytime}}{\log(|\pi_{r_i}|)} = a \cdot \efs + b\) for fitting \(a\) and \(b\).


\subsection{Model for Search Recall }
Next, we model the recall behavior of HNSW index, which uses post-filtering for access control\footnote{Other search techniques (e.g., pre-filtering, single-stage scans) could also be supported in the framework by substituting their respective performance or recall models while preserving the same partitioning logic.}.
(see \S \ref{sec:exp-index}).
Other than the search depth parameter, $\efs$, and the $k$ parameter in the top-$k$ query, the primary factor affecting recall is \emph{access selectivity}: the fraction of documents a user can access in their assigned partitions. For a user $u_i$, access selectivity is defined as the average fraction of accessible documents across all partitions the user needs to query:

\begin{equation}
s_u(u_i) = \frac{1}{|P^{*}(u_i, \Pi)|}\sum_{j \in P^{*}(u_i, \Pi)}\frac{|{auth(u_i)} \cap \pi_j|}{|\pi_j|}
\end{equation}
This is averaged across all users to find the system-wide average access selectivity: $\overline{s_u} = \frac{1}{|U|}\sum_{i=1}^{|U|} s(u_i)$. 
We defined $\overline{s_u}$ using the ``average of ratios" rather than a ``ratio of sums" to better capture multi-partition search dynamics. 
Consider a query spanning Partition A (2,000 documents, 90\% accessible) and Partition B (100,000 documents, 0.5\% accessible). The ``ratio of sums" approach yields 2\% average selectivity, suggesting both partitions require high $ef_s$ values, which would be overkill for Partition A. In contrast, our ``average of ratios" approach yields $\approx 45\%$ average selectivity, which better reflects that relatively small $ef_s$ suffices for high recall: Partition A already has high selectivity, while Partition B's sparse accessible documents limit candidates regardless of $ef_s$.

Our analytical model is based on the observation that recall follows a two-phase pattern as \efs increases: it first grows linearly, then gradually saturates. This leads us to model recall as a piecewise function in Eq~\ref{eq:recalldef}: a linear function for the initial rapid increase and a sigmoid function to capture the saturation effect. 
\vspace{-0.5em}
\begin{equation}
R(\Pi, \efs, \overline{s_u}, k) =
\begin{cases}
\frac{\efs \cdot \overline{s_u}}{k}, & \text{if } \efs \leq \gamma \cdot \frac{k}{\overline{s_u}}, \\
\frac{1}{1 + e^{- \beta \cdot \frac{\overline{s_u}}{k}(\efs - \gamma \cdot \frac{k}{\overline{s_u}})}} + \left(\gamma - \frac{1}{2}\right), & \text{otherwise}.
\end{cases}
\label{eq:recalldef}
\end{equation}
This model connects a partitioning scheme (which determines $\overline{s_u}$) to the search efforts ($\efs$) required to meet the recall target for a \topk query. $\efs$ is mainly governed by the ratio $\frac{k}{\overline{s_u}}$, the expected number of search candidates that must be examined to find $k$ results that pass the permission filter.
The recall model uses two scaling relationships with fitted constants $\gamma$ and $\beta$ with respect to $\frac{k}{\overline{s_u}}$:
\squishitemize
\item \textbf{Transition point} $\gamma \cdot \frac{k}{\overline{s_u}}$: This determines when we transition from linear growth to saturation.
With lower access selectivity, we need to examine more candidates (higher $\efs$) to find $k$ valid results. 
The offset value $\gamma - \frac{1}{2}$ is chosen to ensure continuity of the function value at the transition point.
\item \textbf{Sigmoid steepness} \(\beta \cdot \frac{\overline{s_u}}{k}\): This controls the rate at which recall improves beyond the transition point. Higher access selectivity (\(\overline{s_u}\)) increases the likelihood of retrieving relevant results, resulting in faster recall gains and a steeper curve.
\squishend

To estimate $\beta$ and $\gamma$, we use an RBAC generator to create a permission workload with average access selectivity 0.1. As \efs increases from 1 to 1000 (typical upper limit in databases like pgvector), recall transitions from 0 to 1.  
We execute 1000 random queries across varying \efs values (10 to 1000), collecting average recall per setting. Before each query, we compute access selectivity and retrieve \topk to fit $\beta$ and $\gamma$ in Eq~\ref{eq:recalldef}. 
Our evaluation shows that these parameters are insensitive to the specific workload used for fitting (\S~\ref{sec:eval-model}).

Modeling search performance and recall for HNSW is a challenging problem that has been extensively studied in prior work~\cite{li2020improving, wang2025leafi}. 
Our goal is not to produce perfectly accurate models, but to capture the key factors that influence performance in order to guide \sys’s partitioning decisions.
For this purpose, we make several simplifying assumptions—such as ignoring the effects of data distribution and intrinsic dimensionality on search latency~\cite{hnswdefaults}, and the effects of the correlation between query vectors and user access permissions on search recall~\cite{patel2024acorn}.
Although these simplifications may limit model accuracy when used in isolation, they are sufficient to guide optimization as long as the model can distinguish the relative ranking among candidate partition choices.
Our evaluation shows that \sys’s partitioning algorithm is robust to modeling imperfections (\S~\ref{sec:eval-model}) and works even with asymptotic performance models that require no parameter fitting (\S~\ref{sec:exp-index}).


\section{Dynamic Partitioning Strategy}
Given the analytical models for query performance and recall, we formulate Problem 1 as a constrained optimization problem (formal specification in Appendix A).
This problem is a Mixed-Integer Nonlinear Program (MINLP), which is NP-hard due to its combinatorial search space and nonlinear constraints.
In this section, we introduce \sys’s greedy partitioning strategy that efficiently solves the optimization problem.

\subsection{Greedy Split Algorithm}
\label{sec:greedy}

\begin{figure}[t]
    \centering\includegraphics[width=0.7\linewidth]{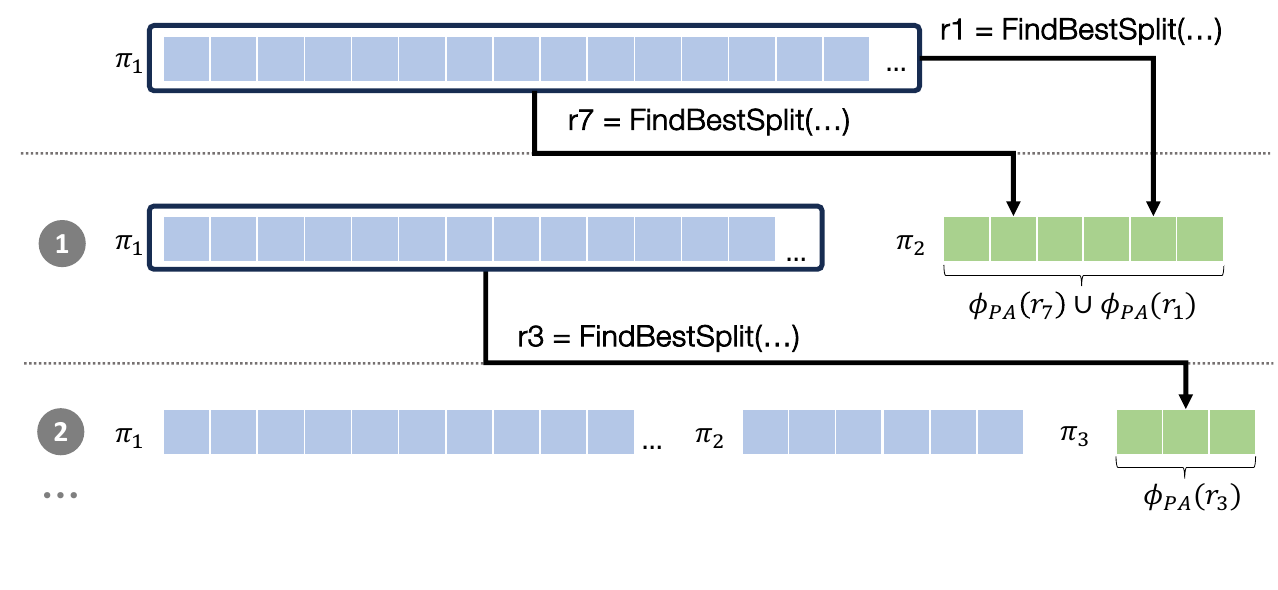}
    \caption{Each iteration of the greedy algorithm, one or more roles are moved from the largest partition to form a new partition (green blocks). The roles are chosen greedily based on the estimate performance improvement using the analytical models.  }
    \label{fig:alg}
\end{figure}
\sys makes a key observation to solve the optimization problem.
The design space is inherently challenging, as it grows exponentially with the total number of documents.
To tackle this scalability challenge, we propose to constrain the search space by ensuring that all documents accessible by a given role are contained in a single partition.

This constraint significantly reduces the search complexity and enables an efficient search algorithm: rather than partitioning over all documents $D$, we only partition over roles $R$, where in typical RBAC settings $|R| \ll |D|$. 
It also provides query latency benefits, since each role’s vector search operates over a single partition of size $|\pi_r|$, yielding logarithmic access cost in $|\pi_r|$.
In contrast, if a role’s documents were split across multiple partitions, the access cost becomes the sum of logarithms across those partitions, effectively growing multiplicatively with the number of partitions (see \S~\ref{sec:analysis} for details).

We propose a greedy partitioning algorithm (Algorithm~\ref{alg:greedy-main}) based on this intuition. The algorithm returns the final partitioning design $\Pi$ and mapping $M$ which maps each partition to its contained roles. $M$ can be used to calculate $P_u(u_i, \Pi)$ as defined in \S~\ref{subsec:problemstatement}. 

As shown in Figure~\ref{fig:alg}, our algorithm follows an iterative splitting approach beginning with a single partition containing all documents and all roles. In each iteration, we identify the largest partition containing more than one role and split it by moving selected roles $r^{*}$ to a new partition. The roles are chosen to: (1) group all documents accessible by $r^{*}$ into the new partition, and (2) maximize query performance improvements based on our analytical model.
Each split introduces a trade-off between memory and query efficiency. Memory increases because documents shared between $r{*}$ and roles remaining in the source partition must be duplicated. However, query latency typically improves as queries involving $r{*}$ benefit from having a higher fraction of accessible documents (increased access selectivity) in the new partition.

\begin{center}
\begin{minipage}{0.85\linewidth}
\begin{algorithm}[H]
\centering 
\algrenewcommand{\algorithmiccomment}[1]{\hfill $\triangleright$ #1}

\caption{\small Greedy Split Algorithm}
\label{alg:greedy-main}

\footnotesize
\begin{algorithmic}[1]
\Statex{\textbf{Input}: Storage constraint $\alpha$}
\Statex{\textbf{Output}: Partitions $\Pi$, partition-to-role mapping $M$ }


\Function{CreatePartition}{$\mathcal{R}$} \Comment{All docs needed by roles in $\mathcal{R}$}
    \State \Return $\bigcup_{r \in \mathcal{R}} \phi_{PA}(r)$ 
\EndFunction

\State $\Pi[1] \gets D$ \Comment{Initialize partition with all documents and roles}
\State $M[1] \gets R$ 

\While{$\sum_{\pi_i \in \Pi} |\pi_i| \leq \alpha |D|$}
    \State $i_{src} \gets$ \Call{FindLargestSplittablePartition}{$\Pi, M$}
    \State $i_{dst} \gets |\Pi|+1$ \Comment{Create new partition}
    \State $\Pi[i_{dst}] \gets \emptyset$, $M[i_{dst}] \gets \emptyset$
    \While{$\sum\limits_{\pi_i \in \Pi} |\pi_i| \leq \alpha |D|$}
        \State $r^{*} \gets$ \Call{FindBestSplit}{$\Pi, M, i_{src}, i_{dst}$}
        
        \State $M[i_{src}] \gets M[i_{src}] \setminus \{r^{*}\}$ \Comment{Update mappings}
        
        \State $M[i_{dst}] \gets M[i_{dst}] \cup \{r^{*}\}$
        
        \State $\Pi[i_{src}] \gets$ \Call{CreatePartition}{$M[i_{src}]$} \Comment{Update partitions}
        \State $\Pi[i_{dst}] \gets$ \Call{CreatePartition}{$M[i_{dst}]$}
        
        \If{$i_{src} \neq$ \Call{FindLargestPartition}{$\Pi, M$} or $r^{*} = \{\}$} 
            \State \textbf{break} 
        \EndIf
    \EndWhile
\EndWhile
\State \Return $\Pi, M$
        \end{algorithmic}
    \end{algorithm}
\end{minipage}
\end{center}

For each round of splitting (line 7, outer loop), the algorithms choose $\Pi[i_{src}]$, the largest partition with more than one role as the source partition for splitting. At the end of the round, the partition gets split into two partition with ids $\Pi[i_{src}]$ and $\Pi[i_{dst}]$. In the inner loop (line 12), the algorithm evaluates the benefit of moving each role from partition $\Pi[i_{src}]$ to the new partition $\Pi[i_{dst}]$, and greedily selects the role $r^{*}$ with the largest improvement of query latency based on the performance model (line 13). 
Then we update $\Pi[i_{dst}]$ by adding documents belonging to $r^{*}$, and update $\Pi[i_{src}]$ by removing documents that are unique to $r^{*}$, as no other roles in $\Pi[i_{src}]$ would have access to these documents. In each iteration of the inner loop, we add one role to $\Pi[i_{dst}]$, until (i) query latency no longer improves, 
(ii) the memory budget would be exceeded, or (iii) $\Pi[i_{\text{src}}]$ is no longer the largest splittable partition. 
Note that $\Pi[i_{\text{dst}}]$ can contain multiple roles at the end of a round.

Algorithm~\ref{alg:evalsplits} implements \textsc{FindBestSplit}, evaluating the cost of each candidate split to determine the most effective one. Two performance models are considered: a role-level model $C_r$ (Eq~\ref{eq:rolecost}) and a user-level model $C_u$ (Eq~\ref{eq:usercost}).
 For $C_r(\Pi)$ or $C_u(\Pi)$, users specify a target recall and input it into the program to compute the corresponding average access selectivity $\overline{s_u}$ for a given partition $\Pi$. Then the \efs is derived using Eq.~\ref{eq:recalldef}.
Intuitively, the role-level model reflects local effects - reduction in per role query time is a good objective that can guide the partitioning from a single partition to the solution of partitioning by roles (one partition per role). The user-level model reflects global effects - reducing user-level cost directly lowers the overall query-time objective.

A split is considered beneficial if $\Delta Q_r < 0$ and $\Delta Q_u$ is below a predefined threshold $\eta$, preventing the greedy algorithm from becoming trapped in locally optimal but globally suboptimal configurations. 
Empirically, if $\Delta Q_r < 0$, it's likely that user-level query time will improve in future splits, even if $ Q_u$ slightly increases in the current iteration.
$\Delta S$ denotes memory cost increase. We normalize the query improvements by the memory cost to identify the split that is most effective per unit memory. 
In practice, $\Delta S$ could also be zero or negative; we then use \((\Delta Q_r + \Delta Q_u)/(\Delta S + \epsilon)\), where a small $\epsilon$ is added to avoid division by zero and ensure numerical stability when $\Delta S \approx 0$, 
prioritizing roles $r$ with $\Delta S < 0$ for reduced memory and query latency.

\begin{center}
\begin{minipage}{0.85\linewidth} 
    \begin{algorithm}[H] 
\caption{\small FindBestSplit Algorithm}
\label{alg:evalsplits}
\footnotesize
\begin{algorithmic}[1]
\Function{FindBestSplit}{$\Pi, M, i_{src}, i_{dst}$}
\State $r^{*} = \{\}$
    \ForAll{$r \in M[i_{src}]$} \Comment{Try each split}
        \State $\pi'_{dst} \gets$ \Call{CreatePartition}{$M[i_{dst}] \cup \{r\}$} 
        \State $\pi'_{src} \gets$ \Call{CreatePartition}{$M[i_{src}] \setminus \{r\}$} 
        
        \State $\Pi' \gets \Pi \setminus \{\Pi[i_{src}], \Pi[i_{dst}]\} \cup \{\pi'_{src}, \pi'_{dst}\}$
        
        \State $\Delta S \gets |\pi'_{src}| + |\pi'_{dst}| - |\Pi[i_{src}]| - |\Pi[i_{dst}]|$ \Comment{$\Delta$Storage}
        \State $\Delta Q_r \gets C_r(\Pi) - C_r(\Pi')$ \Comment{$\Delta$Query time}
        \State $\Delta Q_u \gets C_u(\Pi) - C_u(\Pi')$
        
        \If{$\Delta Q_r < 0$ \textbf{and} $\Delta Q_u < \eta$} \Comment{Check if move is beneficial}
            \If{$(\Delta Q_r + \Delta Q_u) / \Delta S > \Delta_{max}$}
                \State $r^{*} \gets r$
                \State $\Delta_{max} \gets (\Delta Q_r + \Delta Q_u) / \Delta S$
            \EndIf
        \EndIf
    \EndFor
    
    \State \Return $r^{*}$
\EndFunction
        \end{algorithmic}
    \end{algorithm}
\end{minipage}
\end{center}

\subsection{Analysis of the Greedy Algorithm}
\label{sec:analysis}



\minihead{Greedy Algorithm Complexity}
Let $|R|$ be the number of roles and $|D|$ the number of documents.
Alg~\ref{alg:greedy-main} performs at most $|R|$ split operations.
Each split calls \textsc{FindBestSplit} (Algorithm~\ref{alg:evalsplits}), which scans up to $|R|$ roles in the source partition.
For each role, the dominant cost arises from updating the partitions after a hypothetical move (Lines 3-4), where creating a partition from a set of roles requires taking the union of their accessible document sets, giving a worst-case time complexity of $O(|R||D|)$.
Combining these steps yields an overall worst-case time complexity of $O(|R|^3|D|)$.

In practice, the complexity is significantly lower for several reasons.
First, the number of split operations, $|P|$, where $P$ is the set of partitions returned by the algorithm, is primarily determined by the storage constraint $\alpha$ rather than $|R|$, and $|P| \ll |R|$ for realistic storage budgets. Second, the average number of documents accessible per role $|D_{avg}|$ is typically much smaller than $|D|$ (\ie $|D_{avg}| \ll |D|$), reducing the cost of the union to $O(|R||D_{avg}|)$.
In our evaluation on the SIFT10M dataset, the greedy algorithm typically completes in under a minute.
A tighter bound that formally captures the relationship between $\alpha$ and $|P|$, and considers realistic RBAC workload permissions is left for future work.


\minihead{Impact of Splitting Roles Across Partitions} 
For HNSW-based search, the per-partition top-$k$ search cost is 
$O(\efs\log|\pi|)$, 
where $\efs$ is the candidate expansion for the target recall~\cite{patel2024acorn,malkov2018efficient}. 
Under post-filtering semantics, retrieving $k$ accessible results requires $\efs\!\propto\!\tfrac{k}{\overline{s_u}(\pi)}$, 
where $\overline{s_u}(\pi)$ is access selectivity of role $u$ in partition $\pi$.
Therefore, if a role $u$ must access multiple partitions
$P^{\smash{*}}(u,\Pi)$, the total query cost is
\[
\textstyle
\sum_{j\in P^{\smash{*}}(u,\Pi)}
  O\!\left(\frac{k}{\overline{s_u}(\pi_j)} \log|\pi_j|\right).
\]
Splitting a role’s accessible data across multiple partitions introduces overhead, as each partition incurs a separate logarithmic search cost.
When summed across partitions, these costs accumulate as
$\sum_j\log|\pi_j| =\log(\!\prod_j|\pi_j|)$, which can exceed the cost of searching a single combined partition, which scales with $\log(|\bigcup_j \pi_j|)$.
Therefore, splitting a role across multiple partitions generally increase latency, unless each split significantly improves access selectivity (\ie $\overline{s_u}(\pi_j)$) to offset the additional $\log|\pi_j|$ factors.


\subsection{Handling Updates}
\label{sec:update}
We need to consider three cases for how \sys updates its partition scheme under permission workload changes: (1) user insertion/deletion, (2) document insertion/deletion from roles, and (3) role insertion/deletion. Since \sys's partitioning strategy relies on roles, updates can be performed incrementally without full partition rebuilds.
When adding new users to existing roles, \sys identifies the optimal set of partitions for the user and updates user-to-partition routing tables ($P^{*}$, Eq~\ref{eq:trackermin}) accordingly. Conversely, deleting users requires no partition changes; only
the user’s access paths in the routing table are removed.
When inserting new documents into existing roles, \sys locates corresponding partitions and inserts documents.
When deleting documents from roles, \sys removes specific documents from partitions containing the role, but user routing remains unchanged. 
When inserting new roles, the system evaluates performance impact (\(\Delta C/\Delta memory\)), assigns role documents to existing or newly created partitions, and updates routing for users assigned to new roles.
Role deletion involves identifying all partitions containing the role, removing role-exclusive documents, and updating user-to-role assignment $(\phi_{UA})$.
In both cases, only affected partition indices require updates.

While the above procedures handle incremental updates efficiently, significant changes in the permission workload may eventually necessitate full repartitioning to restore balance. For instance, if multiple large roles are added over time, or if role ownership patterns shift dramatically, the initial partition scheme may become suboptimal. However, deciding when repartitioning is necessary represents a substantial research challenge in its own right, as it requires carefully balancing the large upfront cost of repartitioning against potential future query time savings. Prior work in different contexts (e.g., partitioning design for OLAP workloads) has explored reorganization strategies using rule-based heuristics~\cite{snow}, reward functions~\cite{mto} and online algorithms~\cite{rong2024dynamic}. We leave the design of adaptive repartitioning strategies for RBAC workloads with vector data as a promising direction for future work.

\section{Experimental Setup}


\subsection{Dataset and Query Workload}
\label{sec:queryworkload}
We evaluate \sys on three datasets:
(1) Wikipedia-22-12~\cite{huggingfacewikipedia}, from which we sample 1 million entries.
Each entry includes a unique identifier, article title, full text, and a Wikipedia ID (wiki\_id).
We use wiki\_id as the document identifier, and generate 300-dimensional paragraph embeddings using the spaCy model en\_core\_web\_md.
(2) SIFT1M and SIFT10M~\cite{jegou2010productsift}, which are standard vector search benchmarks.
SIFT1M contains 1 million 128-dimensional vectors, while SIFT10M contains 10 million.
Each vector is associated with a unique identifier and a URL pointing to the original image from which the SIFT features were extracted.
We use the unique identifier as the document ID for indexing and retrieval.



Each query $q_i=(u_{q_i}, v_i)$ contains user $u_{q_i}$ and query vector $v_i$. We randomly sample 1000 database vectors as query vectors, and randomly select 1000 user IDs to associate with the queries. By default, each query retrieves the \topk$=10$ nearest vectors within the user's permission, but we also experiment with other $k$ values in the evaluation.
\S~\ref{sec:generators} discusses how we generate RBAC permission workloads with different structures.

We run experiments on a server with Intel i5-13600K processor and 64GB memory. All experiments use PostgreSQL 16 with pgvector 0.8.0. 
Primary evaluations use the HNSW index, with additional experiments applying \sys to ACORN~\cite{patel2024acorn} index.
Experiments use a warm-up procedure to ensure performance measurements reflect in-memory operations only.
We execute each query twice within the same database session: first to warm-up, and second to measure steady-state performance via pure in-memory traversal of the index, eliminating disk I/O and buffer management overhead. We record query selectivity (ratio of user-accessible documents to total documents).


\subsection{RBAC Benchmark}
\label{sec:generators}

Following common practices in RBAC literature, we use permission generators to simulate permissions with different structures for evaluation~\cite{colombo2010rolemining,roleminer,erbac}.
By default, all generators use $|U| = 1000$ users and $|R| = 100$ roles.  

\minihead{Uniform Generator~\cite{roleminer}}
This generator produces permission data without specific structure. Two parameters are required: maximum roles per user $(m_r)$ and maximum documents per role $(m_p)$. Both role-permission assignment ($\phi_{PA}$) and user-role assignment ($\phi_{UA}$) are uniformly generated~at~random.


We evaluate performance using two different sets of parameters. 
Performance evaluation uses two parameter sets. Uniform-$\alpha$, employed in \S~\ref{sec:main-results}: $m_r$ = 2, $m_p$ = $|D| / |R| \times 5$. Uniform-$s$, employed in \S~\ref{sec:sensitivity-analysis}: $m_r$ = 1, $m_p$ = $|D| / |R| \times 9$.

\begin{table}[t]
    \centering
    \caption{Comparison of workload configurations. }
    \small  \renewcommand{\arraystretch}{0.9} 
    \small 
    \begin{tabular}{ l c c c c }
        \toprule
        & \textbf{Tree-$\alpha$}& \textbf{Uniform-$\alpha$} & \textbf{ERBAC-$\alpha$}  & \textbf{ERBAC-$\beta$} \\ 
        \midrule
        Avg Access Selectivity & 0.036 & 0.054 & 0.128  & 0.285 \\
         Max Roles Per User  & 1 & 3 & 3 & 9 \\
        \midrule
        Role Partition Memory Overhead & 3.5$\times$ & 3.8$\times$ & 7.0$\times$ & 7.0$\times$ \\
        User Partition Memory Overhead & 3.5$\times$ & 74.9$\times$ & 134$\times$ & 408$\times$ \\
        \bottomrule
    \end{tabular}
    \label{tab:erbac-comparison}
\end{table}

\minihead{Tree-based Generator~\cite{colombo2010rolemining}}
This generator models organizational hierarchical role structures (e.g., CEO $\rightarrow$ department heads $\rightarrow$ team leads $\rightarrow$ employees).
Parameters include: tree height ($h$), lower-bound ($b_0$) and upper-bound ($b_1$) for children per internal node. 
\squishitemize
\item \emph{Tree and role construction}: Generate tree \( T \) with height \( h \), constructing role hierarchy recursively from root, assigning random children in range \( [b_0, b_1] \) until role pool \( |R| \) exhausted or height \( h \) reached. Each tree node represents an organizational role in hierarchical structure.
For instance, consider a company with multiple departments, each containing several offices; the root node grants universal permissions, while a leaf node might represent a specialized role in a specific office.
\item \emph{Role-permission assignment ($\phi_{PA}$):} 
Split document set $D$ into $|R|$ subsets, assigning each to its corresponding role. Roles inherit all permissions from ancestor roles, so each role $r$'s effective document access is the union of documents directly assigned to $r$ and its ancestors\footnote{For example, an employee in the Business Office of Department A would have access to company-wide permissions (from the root), Department A-specific permissions (from an intermediate node), and permissions unique to the Business Office (from a leaf node). Department-wide permissions are exclusive to employees in that department and never shared across departments, while office-specific permissions are restricted to employees of that office and not shared with other offices.}.
\item \emph{User-role assignment ($\phi_{UA}$):} Evenly distribute the set of users across all roles in the tree, excluding the root role. Each user is assigned to one role. 
\squishend

We use two parameter sets. Tree-$\alpha$ (used in \ref{sec:main-results}): $h = 4$, $b_0 = 3$, $b_1 = 4$. Tree-$s$ shares Tree-$\alpha$ parameters but uses Poisson distribution for $\phi_{PA}$, adjusting parameters to vary~permission~selectivity.

\minihead{ERBAC Generator~\cite{erbac,colombo2010rolemining}}
\label{sec:erbac-generator}
Based on Enterprise Role-Based Access Control (ERBAC) model using two-level layered role hierarchy commonly found in real-world organizations. Introduces functional roles (define job functions, hold direct permissions) and business roles (group functional roles, inherit their permissions). Notably, business roles represent actual roles (\( R \)) assigned to users (\( U \)). This generator requires five parameters: the number of functional roles ($n_{fr}$), the number of business roles ($n_{br}$), the maximum number of permissions a functional role can have ($m_p$), the maximum number of functional roles a business role can connect to ($m_{fr}$), and the maximum number of business roles a user can have ($m_{br}$). 
\squishitemize
    \item \emph{Generate functional and business roles:} For each functional role $r$, randomly select the number of permissions $m(r)$ between 1 and the total number of permissions $|D|$. Assign $m(r)$ randomly chosen permissions from the set $D$ to $r$. For each business role $r$, randomly select the number of functional roles $m(r)$ from $\{1, \dots, m_{fr}\}$ and randomly assign $m(r)$ functional roles to $r$. For each business role, the permission set is the union of all permissions held by its associated~functional~roles.
    \item \emph{Assign business roles to users:} For each user $u$, randomly select the number of business roles $m(u)$ from $\{1, 2, \dots, m_{br}\}$. Assign $m(u)$ randomly chosen business roles to $u$. For each user the permission is the union of all permissions inherited from the assigned business roles.
\squishend

We evaluate performance using two different sets of parameters. 
The first set, ERBAC-$\alpha$, used in \ref{sec:main-results}, is defined as follows: $n_{fr} = 40$, $n_{br} = 100$, $m_{fr} = 3$, $m_{br} = 3$, $n_p = |D|$, and $m_p = |D| / 25$. The second set, ERBAC-$\beta$, also used in \ref{sec:main-results}, is similar to ERBAC-$\alpha$, except that $m_{br} = 9$. The third set, ERBAC-$s$, used in \ref{sec:sensitivity-analysis}, shares the same basic parameters as ERBAC-$\alpha$, with the exception of $m_{br} = 1$.

\subsection{Baseline Methods and Metrics}
We compare \sys with four baselines: 
\squishitemize
\item \textbf{Role Partition} creates separate partition per role, which is efficient for single-role users but may underperform user partitioning for multi-role users due to the need to aggregate documents from multiple partitions during query execution.
\item \textbf{User Partition} creates separate partition per unique role combination, so that users with the same roles share one partition.
\item \textbf{Row-level Security (RLS)} applies PostgreSQL's row-level security feature to filter query results based on permissions after performing similarity search, serving as an example of post-filter methods.
\item \textbf{HQI}~\cite{hqi} extends QD-tree to build partitions based on frequent predicates observed in a query workload, serving as an example of partition-based methods for hybrid vector search. However, it only generates non-overlaping partitions.  
\squishend

For all methods, we initialize \efs from the analytical model’s estimate and adjust it by up to 30–40\% through a small parameter sweep. We then use binary search to fine-tune the value, which typically converges within 1–2 iterations. 

The methods are compared using the following metrics: 
\squishitemize
\item \textbf{Memory}: total index footprint in memory normalized with respect to memory consumption for Row-Level Security (RLS).
The index size is calculated using PostgreSQL's pg\_indexes\_size(), and scales roughly linearly with the number of vectors.
\item \textbf{Query Latency}: the average execution time per query, measured only on the second run of each query (the first run warms up the cache and is excluded from timing).
\item \textbf{Recall@k}: the fraction of true \topk nearest neighbors (from ground truth) that appear in the retrieved \topk results, measuring retrieval accuracy.
\squishend

\section{Evaluation}
We evaluate \sys using the described permission generators and datasets. Our experiments~show:
\squishitemize
    \item 
   \sys enables efficient trade-off between query latency and memory overhead, achieving up to 6$\times$ speedup with 1.4$\times$ memory at \topk=10, and up to 13.5$\times$ speedup with 1.24$\times$ memory at \topk=100 compared to RLS, while achieving comparable query performance to Role Partition with 84\% and 90.4\% reduction in memory overhead respectively (\S~\ref{sec:main-results}, \S~\ref{sec:sensitivity-analysis}). 
    
    \item 
    \sys complements hybrid vector search indices such as ACORN, achieving up to 3.5$\times$ speedup at 1.2$\times$ memory compared to using ACORN alone (\S~\ref{sec:exp-index}). 
    \item \sys’s partitioning algorithm is robust to small variations in performance model parameters (\S~\ref{sec:eval-model}), and physical partitioning consistently provides a better latency–memory trade-off than logical partitioning (\S~\ref{sec:eval-logical}).
    \item \sys’s benefits increase with larger \topk values, lower average access selectivity, and hierarchical role structures (\S~\ref{sec:sensitivity-analysis}), and it can be efficiently maintained under updates to permission workloads (\S~\ref{sec:updateeval}).

    
\squishend

\begin{figure*}[htbp]
  \centering
  \begin{minipage}{\linewidth}
        \centering
        \includegraphics[width=0.6 \linewidth]{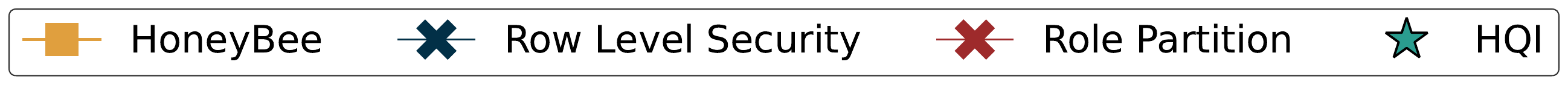}
    \end{minipage}
  \begin{subfigure}{0.24\linewidth}
    \centering
    \includegraphics[width=\linewidth]{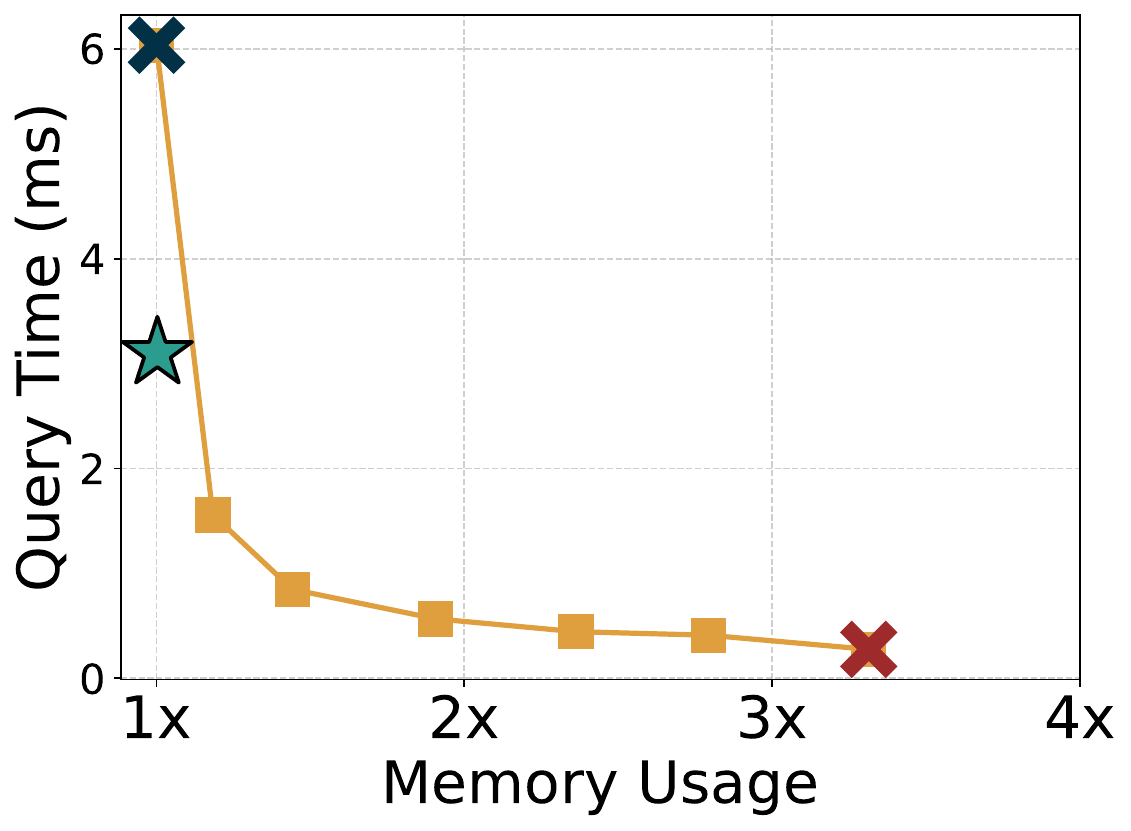}
    \caption{Tree - $\alpha$}
    \Description{A graph of query time versus \Revision{memory} for a Treebased workload, showing the impact of hierarchical role structures.}
    \label{fig:query-time-vs-storage-treebased}
  \end{subfigure}
  \hfill
  \begin{subfigure}{0.24\linewidth}
    \centering
    \includegraphics[width=\linewidth]{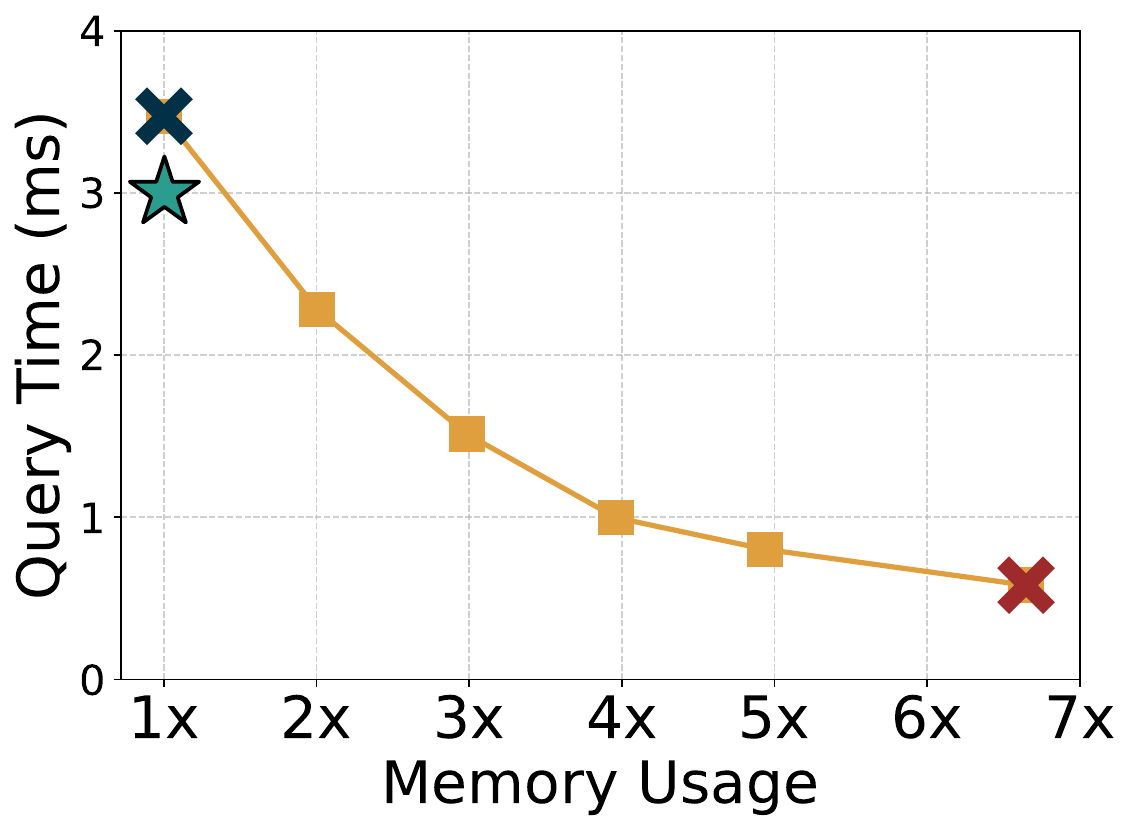}
    \caption{ERBAC - $\alpha$}
    \Description{A plot showing query time versus \Revision{memory} for ERBAC - $\alpha$, demonstrating the trade-offs in different workload configurations.}
    \label{fig:query-time-vs-storage-erbac-alpha}
  \end{subfigure}
  \hfill
  \begin{subfigure}{0.24\linewidth}
    \centering
    \includegraphics[width=\linewidth]{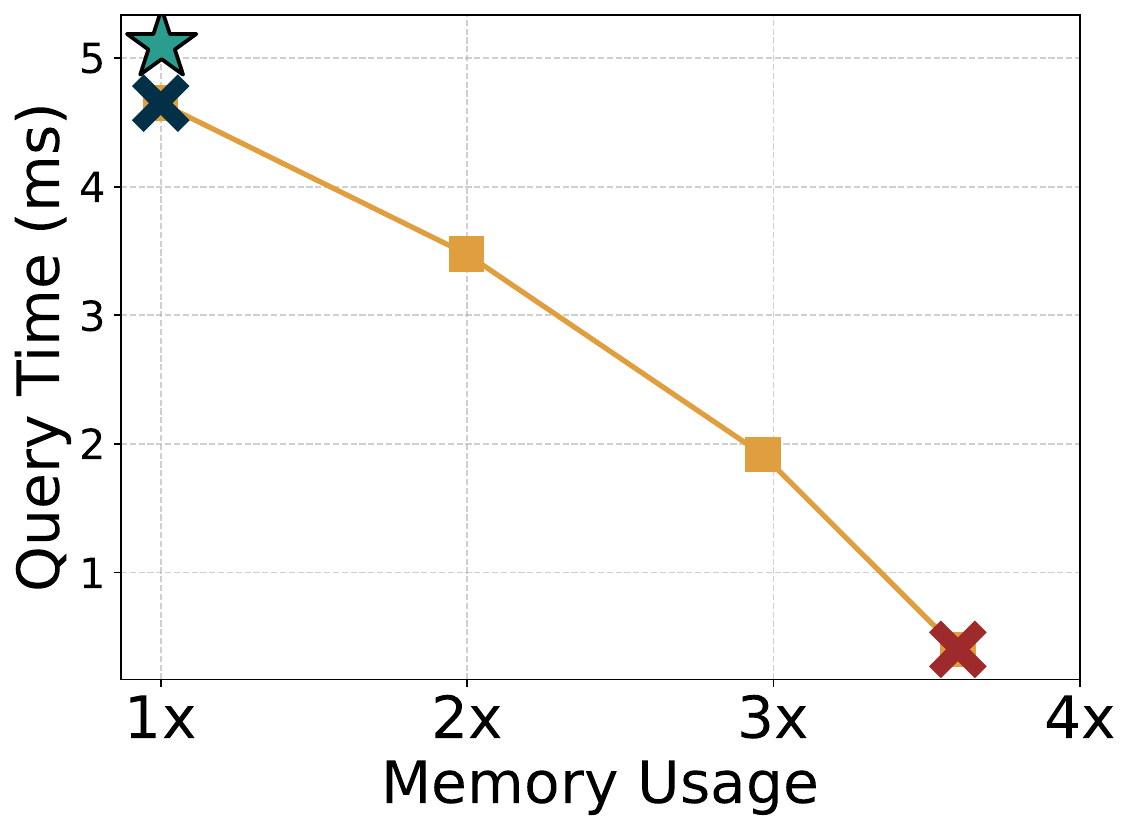}
    \caption{Uniform - $\alpha$}
    \Description{A graph representing query time versus memory for a random workload distribution, used as a baseline for comparison.}
    \label{fig:query-time-vs-storage-random}
  \end{subfigure}
    \hfill
  \begin{subfigure}{0.24\linewidth}
    \centering
    \includegraphics[width=\linewidth]{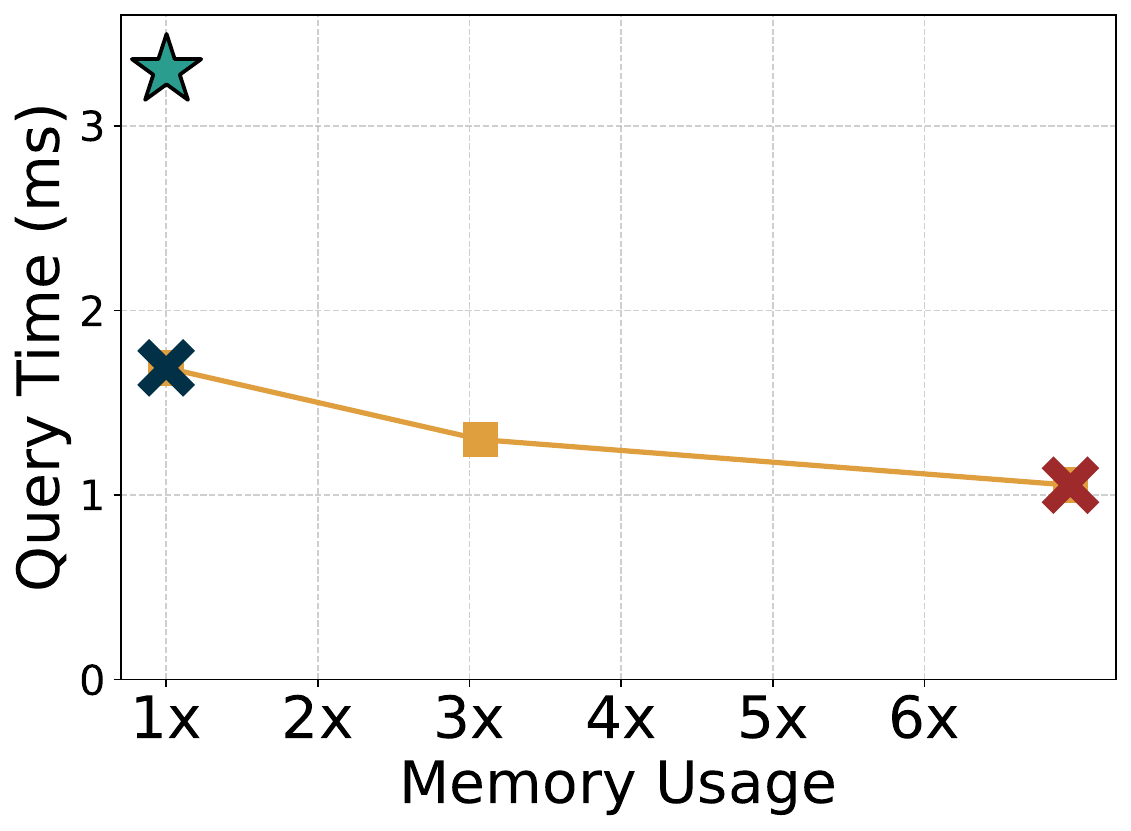}
    \caption{ERBAC - $\beta$}
    \Description{A plot showing query time versus \Revision{memory} for ERBAC-$\beta$, highlighting how query latency is affected by memory constraints.}
    \label{fig:query-time-vs-storage-erbac-beta}
  \end{subfigure}
      \vspace{-0.5em}
\caption{Trade-off between Query Time and Memory Usage across Permission Workloads.}
\Description{A set of four subfigures comparing query time versus \Revision{memory} across different permission workload types: ERBAC, ERBAC-$\beta$, Random, and Tree. }
\label{fig:query-time-vs-storage}
\end{figure*}

\begin{figure*}[htbp]
  \centering
    \begin{minipage}{\linewidth}
        \centering
        \includegraphics[width=0.6\linewidth]{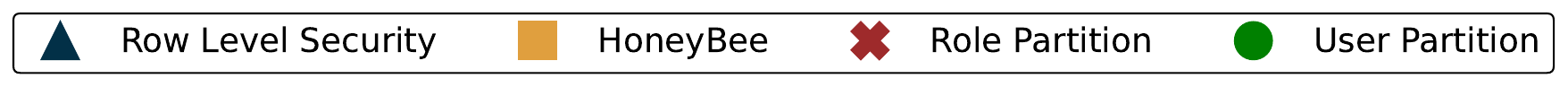}
    \end{minipage}
    \begin{subfigure}{0.24\linewidth}
    \centering
    \includegraphics[width=\linewidth]{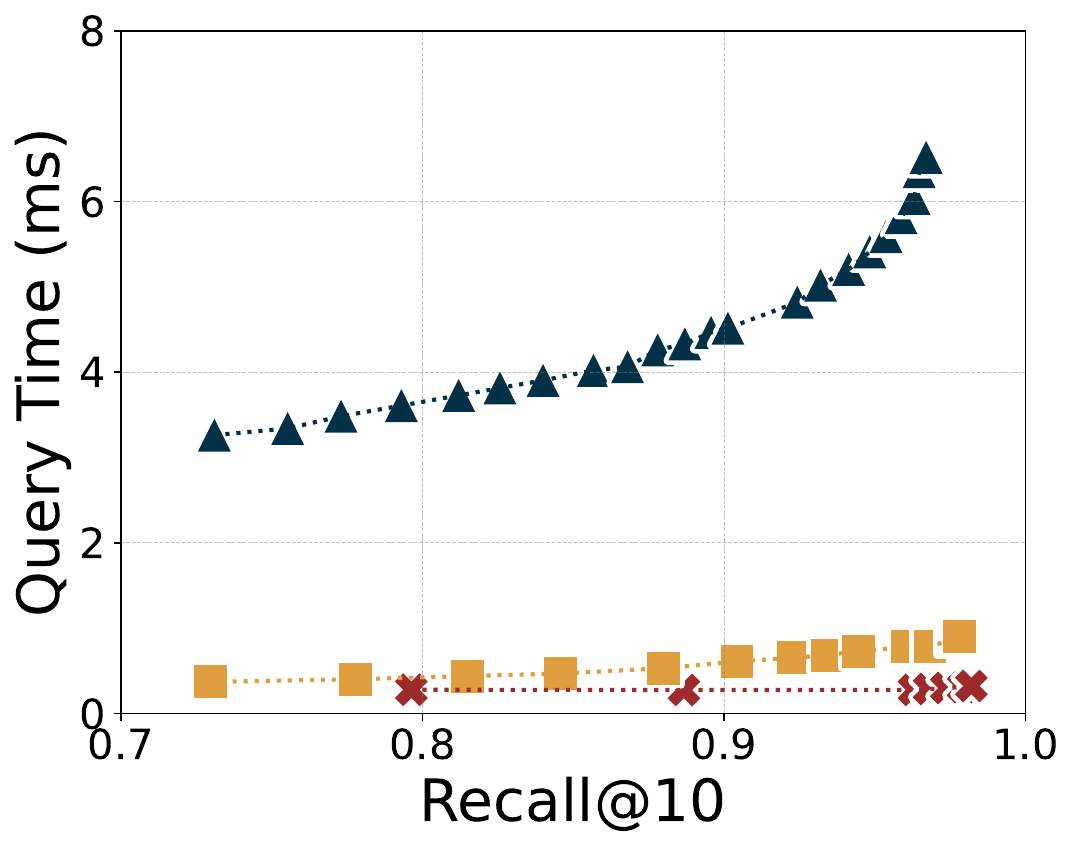}
    \caption{Tree - $\alpha$}
    \Description{A figure showing query time versus recall for a Tree workload, highlighting hierarchical permission structures.}
    \label{fig:qps-recall-treebase}
  \end{subfigure}
    \hfill
    \begin{subfigure}{0.24\linewidth}
    \centering
    \includegraphics[width=\linewidth]{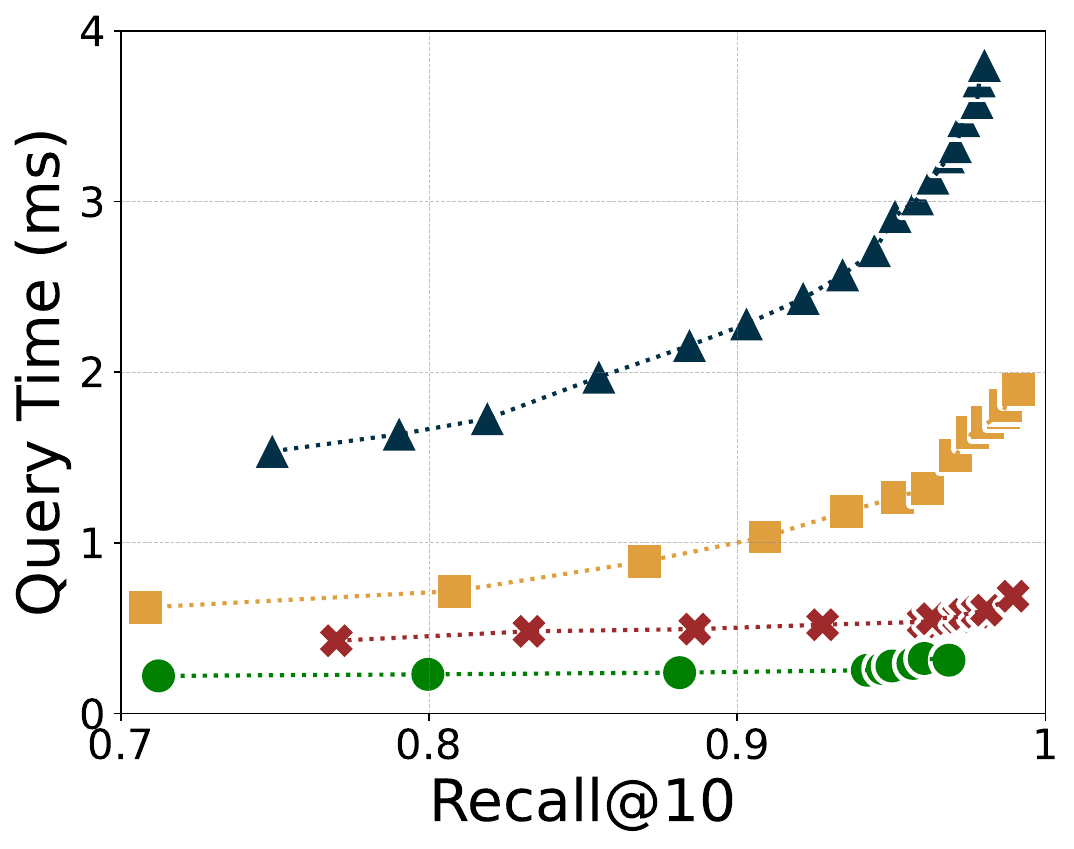}
    \caption{ERBAC - $\alpha$}
    \Description{A plot showing the relationship between query time and recall for ERBAC, demonstrating how increasing recall affects latency.}
    \label{fig:qps-recall-erbac-alpha}
  \end{subfigure}
  \hfill
  \begin{subfigure}{0.24\linewidth}
    \centering
    \includegraphics[width=\linewidth]{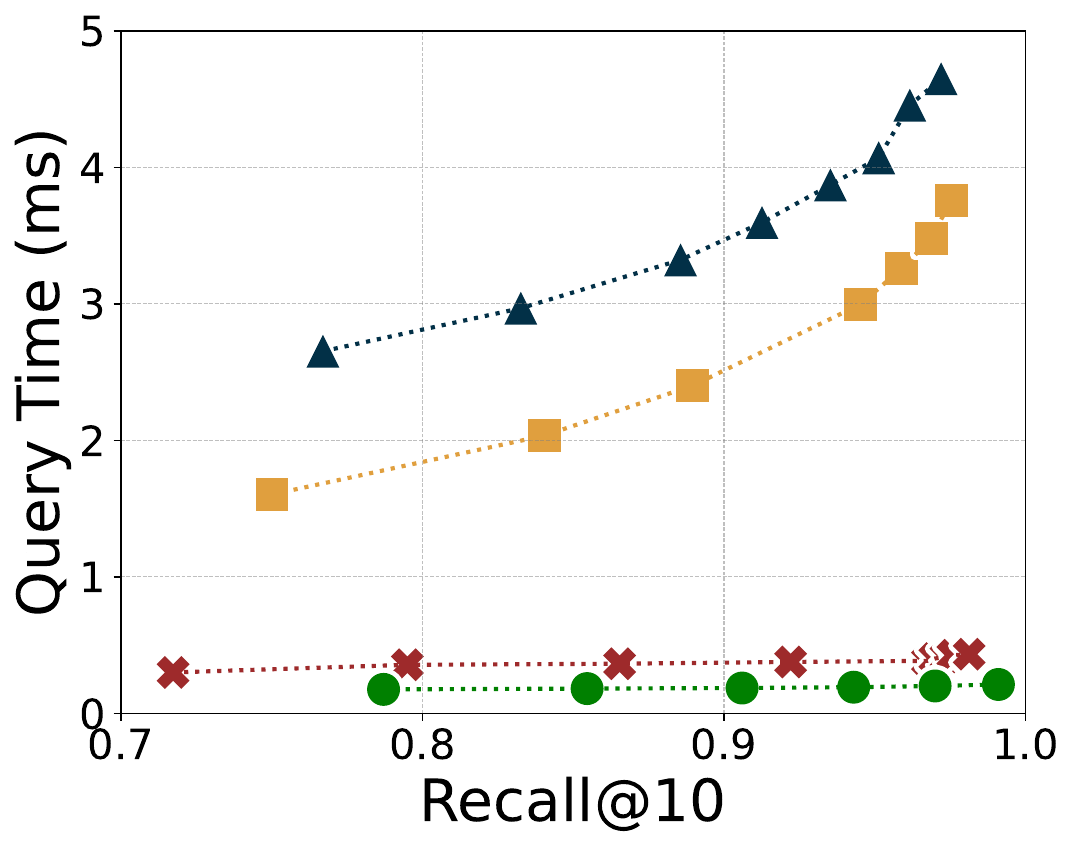}
    \caption{Uniform - $\alpha$}
    \Description{A query time versus recall plot for a randomly generated workload, used as a baseline for performance evaluation.}
    \label{fig:qps-recall-random}
  \end{subfigure}
  \hfill
    \begin{subfigure}{0.24\linewidth}
    \centering
    \includegraphics[width=\linewidth]{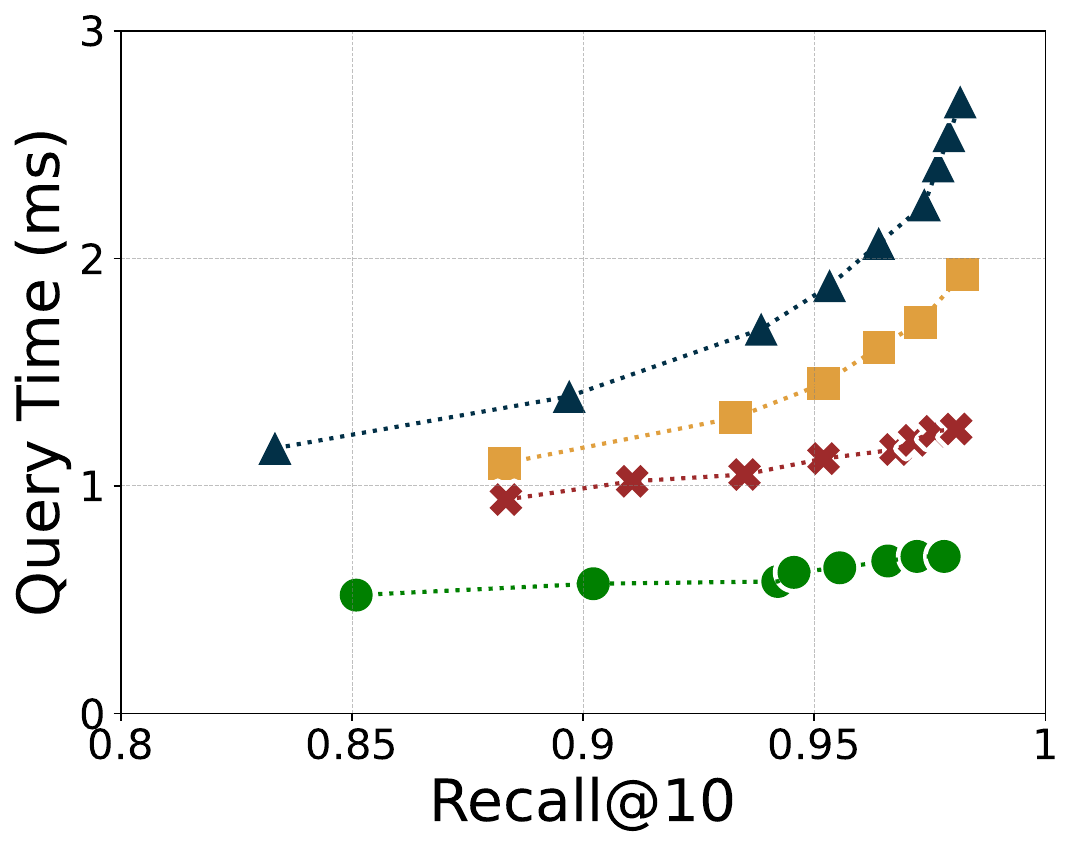}
    \caption{ERBAC-$\beta$}
    \Description{A plot of query time versus recall for ERBAC-$\beta$, illustrating the impact of \rthree{}{access selectivity} on retrieval performance.}
    \label{fig:qps-recall-erbac-beta}
  \end{subfigure}
  \vspace{-0.5em}
\caption{Query Time vs Recall under RLS and Different Partitioning Strategies. For \sys we used the following memory points for the four workloads: (a) 1.4$\times$; (b) 3.0$\times$; (c) 1.9$\times$; (d) 3.2$\times$.}
\Description{A set of four subfigures comparing query time versus recall across different partitioning strategies, including ERBAC, ERBAC-$\beta$, Random, and Tree.}
\label{fig:qps-recall-full-analysis}
\end{figure*}

\subsection{Trade-off between Memory, Latency, Recall }
\label{sec:main-results}

In this section, we evaluate the trade-off between memory overhead, query latency, and recall under different permission workloads.

\autoref{fig:query-time-vs-storage} shows the trade-off between memory overhead and average query latency across different permission workloads while maintaining 0.95 recall. RLS (navy cross) offers minimal memory ($1\times$)
but high query latency, while Role Partition (red cross) delivers low query latency at significantly higher memory cost. 
\sys occupies the intermediate space, producing partitioning strategies ranging from 1 to \( |R| \) partitions based on the memory constraint (\eg \sys generates 20 partitions for Tree-$\alpha$ with 1.4$\times$ memory constraint). 
Curves closer to the bottom-left corner indicate better latency–memory trade-offs.. 
Note that \sys's greedy algorithm \emph{estimates} space needed as total documents in all partitions rather than measuring actual partition sizes, and permits exceeding the limit after the last partitioning step. This may cause minor deviation from the target space factor ($\alpha$) when measured using actual partition sizes, though the violation remained within 6\% in all experiments.

We also compare against HQI~\cite{hqi}, another partition-based method that builds separate indices for each partition.
It recursively splits data according to query predicates until each partition falls below a minimum size, which we set to $(\text{number of vectors}) / (\text{number of roles})$ to align with Role Partition.  
Like RLS, HQI produces a single partitioning at 1$\times$-memory, appearing as isolated points in \autoref{fig:query-time-vs-storage}.
Overall HQI shows unstable performance: it outperforms RLS on some workloads (\eg, Tree-$\alpha$: $\mathbf{2.0\times}$ faster, ERBAC-$\alpha$: $\mathbf{1.09\times}$ faster) but lags on others (\eg, Uniform-$\alpha$: $\mathbf{1.1\times}$ slower, ERBAC-$\beta$: $\mathbf{1.8\times}$ slower). 
This behavior is expected because HQI was designed for mostly non-overlapping predicates (\eg single-valued categorical attributes, numeric range queries), where its splits naturally create disjoint partitions aligned with queries.
In contrast, in RBAC workloads, documents can belong to multiple roles, and HQI recursively splits partitions using predicates for different roles.
As a result, the documents needed for a single role are spread across multiple small, disjoint partitions, forcing each query to access many partitions and increasing overhead. Without a memory–latency–aware model, these heuristic splits lead to unstable query performance under such many-to-many access patterns.

\autoref{fig:qps-recall-full-analysis} illustrates the relationship between query recall and latency under fixed memory constraints. We vary the $\efs$ to tune this trade-off. User Partition (green curve) consistently delivers lowest query latency, followed by Role Partition (red curve), while RLS delivers highest latency (black curve). However, \autoref{tab:erbac-comparison} shows this performance requires substantial memory costs, with User Partition requiring 74-408$\times$ RLS memory, while Role Partition typically requires up to 10$\times$ memory. The minimal latency difference between these approaches suggests limited benefits from extending memory beyond Role Partition.
\sys (yellow curve) produces partition strategies between Role Partition and RLS, showing slower growth in query latency compared to RLS at~high~recall~levels.

We analyze performance differences across workloads below:

\noindent\textbf{Tree$-\alpha$.} As shown in \autoref{fig:query-time-vs-storage-treebased}, \sys achieves a $6\times$ latency reduction over RLS at $1.4\times$ memory, approaching Role Partition performance with 84\% less memory.
\autoref{fig:qps-recall-treebase} further shows that at $1.4\times$ memory, \sys significantly outperforms RLS and closely matches Role Partition at $3.5\times$ memory across recall levels. Moreover, \sys shows a much slower growth in query performance as recall improves compared to RLS. User User Partition is omitted since it matches Role Partition performance under one-role-per-user mappings.
Overall, \sys offers a highly effective memory and latency trade-off in this workload.

\noindent\textbf{ERBAC-$\alpha$.} \sys achieves a smooth, convex trade-off curve between RLS and Role Partition (\autoref{fig:query-time-vs-storage-erbac-alpha}), indicating effective trade-off.
At approximately 3$\times$ memory, query latency exceeds twice RLS speed. 
\autoref{fig:qps-recall-erbac-alpha} shows that User Partition achieves slightly lower latency but at $134\times$ the memory of RLS, compared to $7\times$ for Role Partition, suggesting diminishing returns beyond moderate memory budgets.

\noindent\textbf{ERBAC-$\beta$.} This workload assigns up to 9 roles per user (versus 3 in ERBAC-$\alpha$), creating higher user access selectivity where post-filtering approaches like RLS perform well. 
As a result, RLS performs comparably to Role Partition despite a $6\times$ memory gap (\autoref{fig:query-time-vs-storage-erbac-beta}).
At $1\times$ memory, \sys converges to the RLS configuration, while at $3\times$ memory \sys produces solution with query latency between RLS and Role Partition (\autoref{fig:qps-recall-erbac-beta}).
User Partition remains the fastest but consumes up to $408\times$ RLS memory.

\noindent\textbf{Uniform.} \sys shows less benefit trading off memory or query latency versus previous workloads, seen by concave shape on intermediate points in \autoref{fig:query-time-vs-storage-random}.
At 1.9$\times$ memory, latency improves only 1.3$\times$ over RLS, remaining far from Role Partition performance at around $3.5\times$ memory. 
\autoref{fig:qps-recall-random} shows similar recall–latency growth trends for both \sys and RLS.
Random role–permission assignments limit the separability of permission subsets, reducing \sys’s effectiveness.
\S~\ref{sec:sensitivity-analysis} provides additional analysis on the impact of permission workload structure and access selectivity.

\subsection{Evaluation with Hybrid Search Index}
\label{sec:exp-index}
\begin{figure}[t]
  \centering
    \begin{minipage}{0.7\linewidth}  
    \centering
  \begin{subfigure}[b]{0.48\linewidth}
    \centering
    \includegraphics[width=\linewidth]{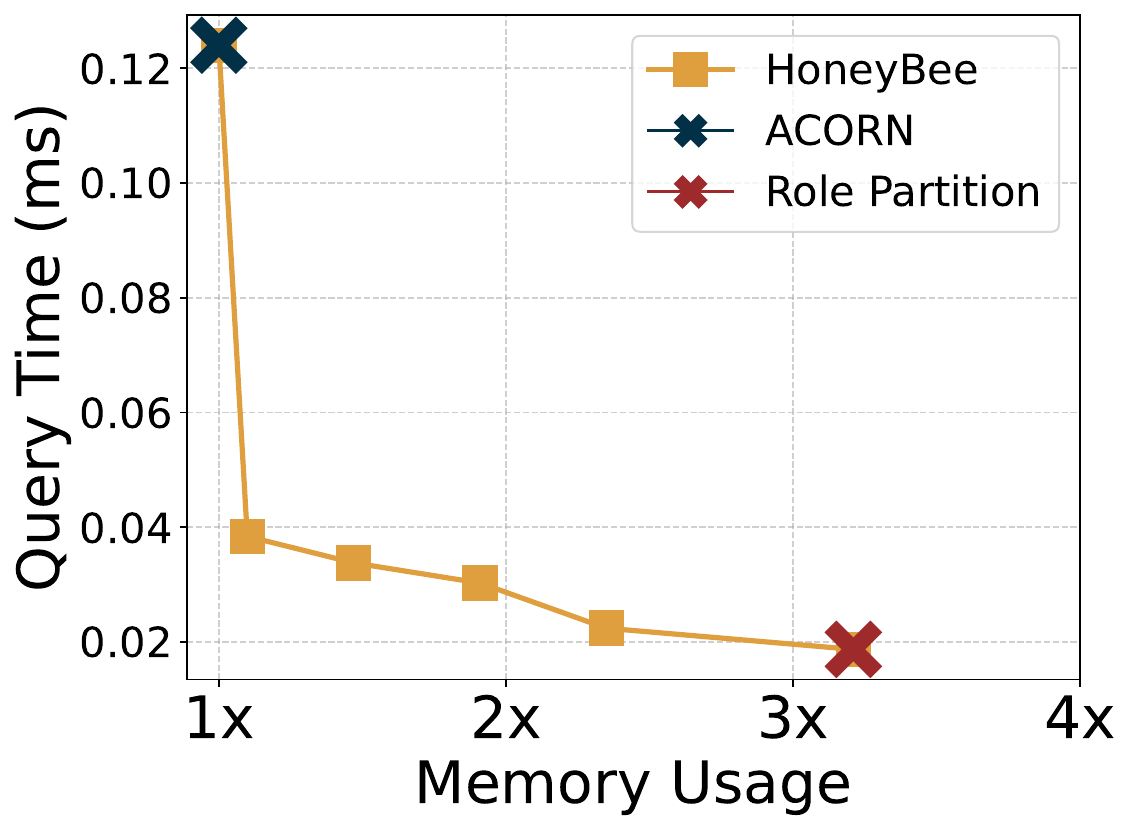}
    \caption{Query Time vs Memory Usage}
    \label{fig:query-time-vs-storage-acorn-treebased}
  \end{subfigure}
  \hfill
  \begin{subfigure}[b]{0.46\linewidth}
    \centering
    \includegraphics[width=\linewidth]{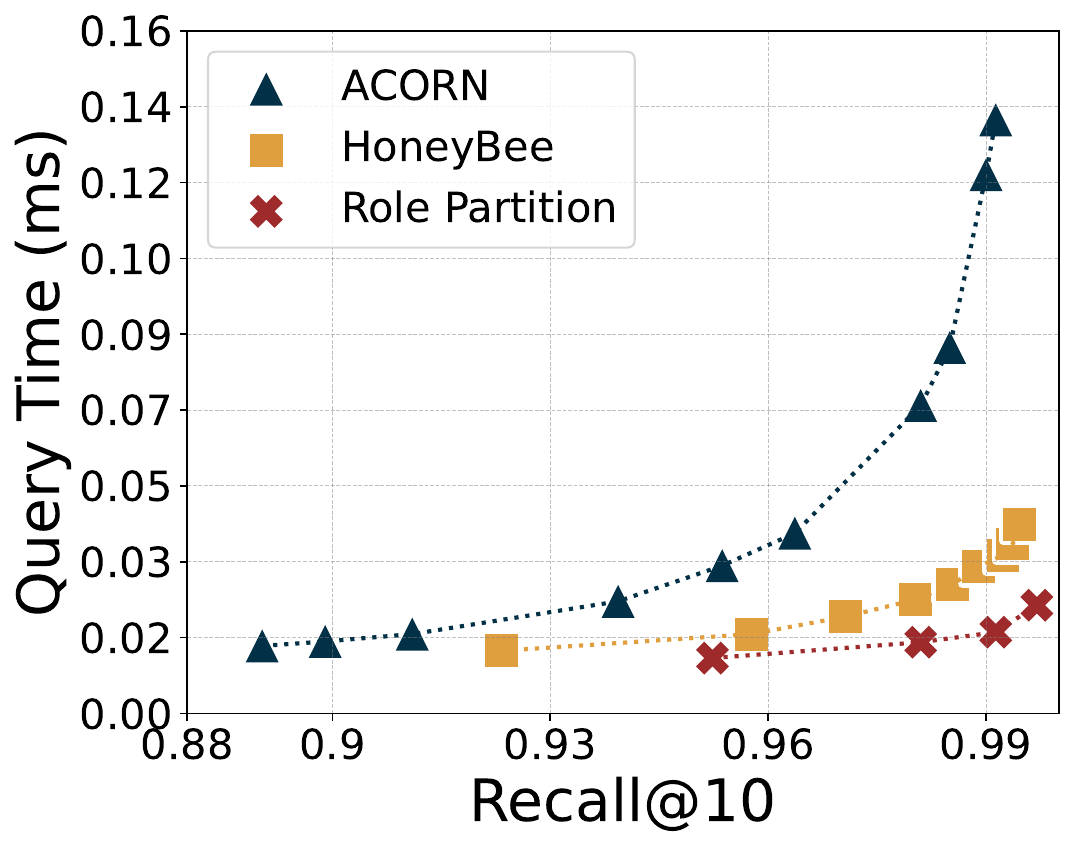}
    \caption{Query Time vs Recall}
    \label{fig:qps-recall-acorn-treebase}
  \end{subfigure}

\end{minipage}
  \vspace{-0.5em}
\caption{Performance of \sys using the ACORN index using the Tree-$\alpha$ workload.}
  \label{fig:acorn-treebased-performance}
\end{figure}

To demonstrate \sys compatibility with hybrid search indices, we replace the default HNSW index with ACORN~\cite{patel2024acorn}, a state-of-the-art index built on HNSW for hybrid vector search.

\sys uses an \textbf{ACORN-based performance model} to guide partitioning.
ACORN provides an analytical expression of search cost:
$O\big((d+\gamma)\cdot \log(s\!\cdot\! n) + \log(1/s)\big)$,
where $d$ is the vector dimension, $s$ is predicate selectivity, $n$ is dataset size, and $\gamma$ is the neighbor-expansion factor.
We use this expression directly as ACORN's search performance model \emph{without any parameter fitting}.
Compared to the HNSW-based model, it preserves the logarithmic dependence on $n$ while introducing ACORN-specific terms, especially $\gamma$, which controls tradeoff between graph connectivity and filtering efficiency.
As a result of ACORN's index design,  it can maintain high recall even under low selectivity, so no separate recall model is required.


For each partition, we apply ACORN if permission filtering is required and HNSW otherwise (e.g., partition containing a single user). Using Tree-$\alpha$ generator produces low-selectivity settings ($\approx$0.03) where ACORN significantly outperforms post-filtering approaches like RLS.
The experiment uses ACORN's implementation on the Faiss library with PostgreSQL. \sys creates partitions in PostgreSQL. Then, ACORN indexes are built on the document table (the default single table with all documents) and its partitions. We disable multi-threading and caching for fair comparison.

\autoref{fig:query-time-vs-storage-acorn-treebased} shows at 1.2$\times$ memory, \sys is roughly 3.5$\times$ faster than using a single ACORN index on all documents (1$\times$ memory).
\autoref{fig:qps-recall-acorn-treebase} show that both \sys and Role Partition maintain stable latency as recall increases, while a single ACORN index experiences a significant increase in latency at higher recall levels.
Overall, these results demonstrate that as a partition-based method, \sys is synergestic with index-based hybrid search methods like ACORN -- by strategically replicating vectors across different indices, \sys can further improve query latency compared to using a single hybrid index on the entire dataset.

\subsection{Evaluation of Performance Model}
\label{sec:eval-model}
\begin{figure}[t]
  \centering
    \begin{minipage}{0.7\linewidth}  
    \centering
  \begin{subfigure}{0.47\linewidth}    \includegraphics[width=\linewidth]{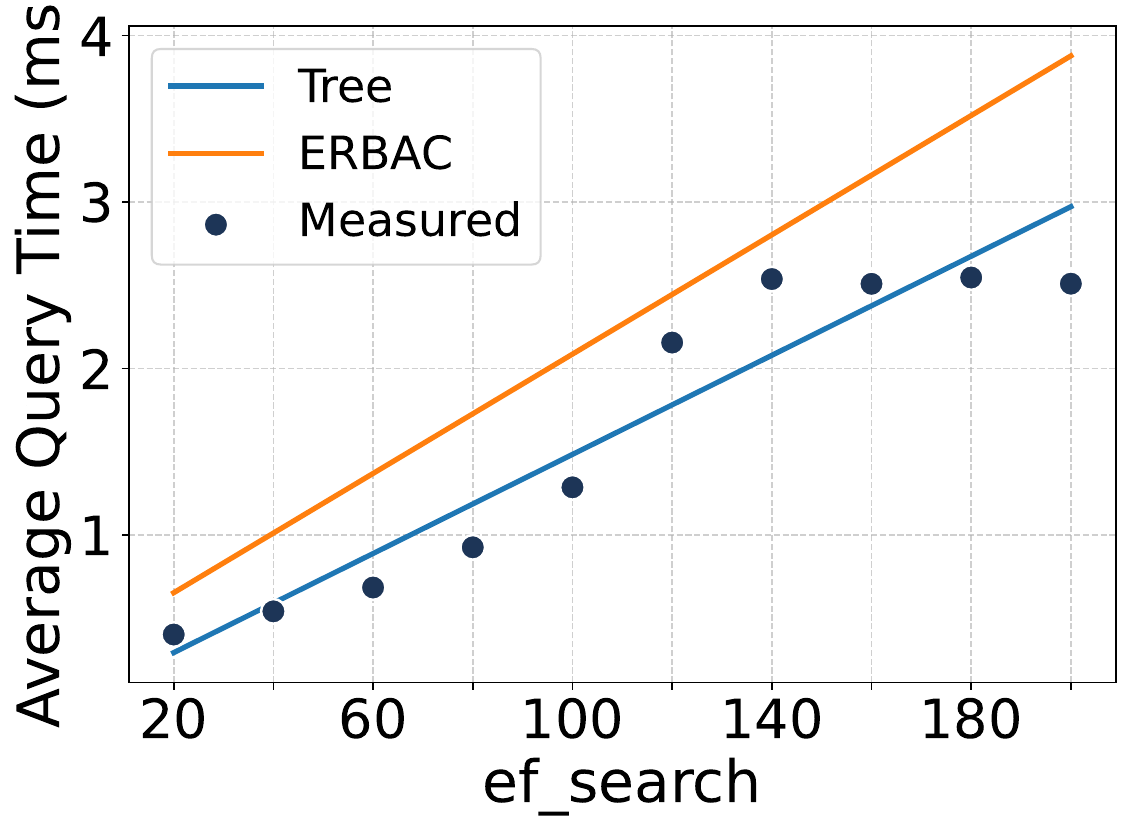}
    \caption{Query Model}
    \label{fig:query-validation-tree}
  \end{subfigure}
  \hfill
  \begin{subfigure}{0.47\linewidth}
    \includegraphics[width=\linewidth]{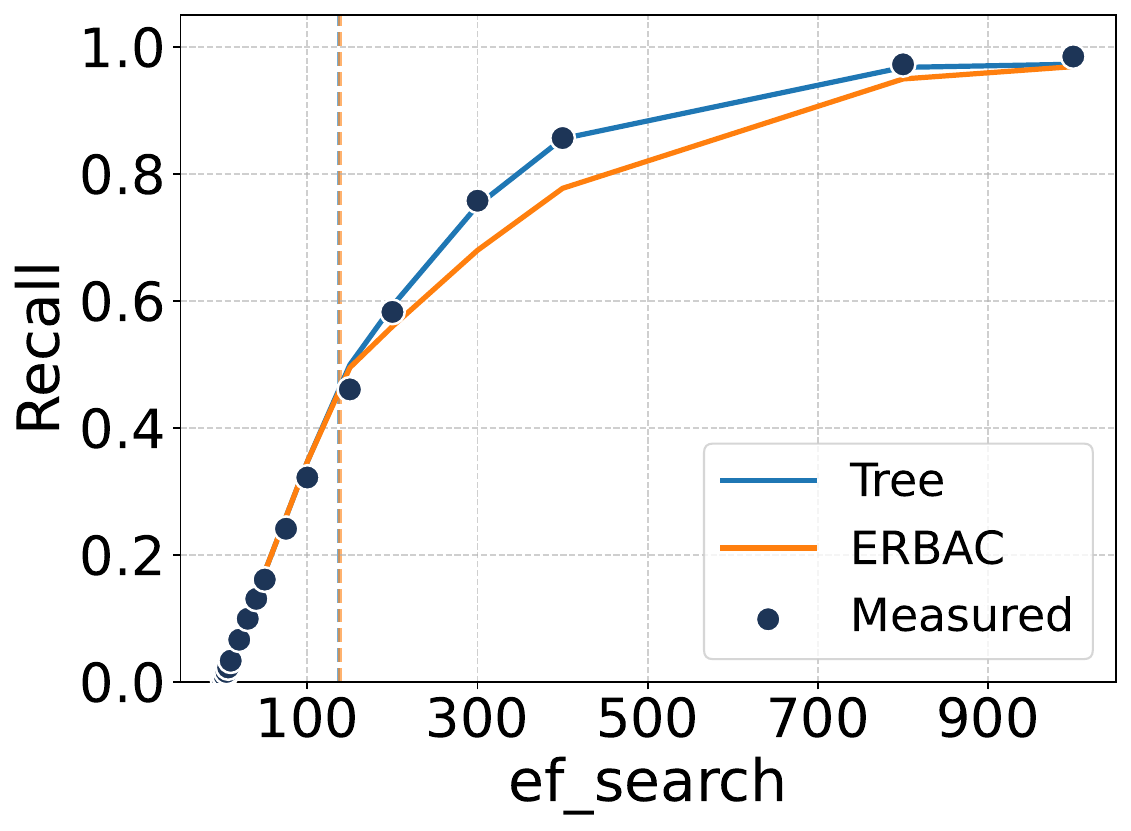}
    \caption{Recall Model}
    \label{fig:recall-validation-tree}
  \end{subfigure}
\end{minipage}
  \vspace{-0.5em}
  \caption{Model validation on the Tree workload using performance parameters obtained from Tree (left) and ERBAC (right) models.}
  \label{fig:model-validation}
\end{figure}

We evaluate the accuracy of \sys's performance models and their sensitivity to the workload used for fitting.
We compare two workloads \textit{Tree-$\alpha$} (one user–one role) and \textit{ERBAC-$\alpha$} (one user–many roles).
Model parameters are firstly fitted on each workload separately, then validated on \textit{Tree-$\alpha$} by comparing \textbf{estimated} latency and recall with \textbf{actual} measurements.
As shown in Figures~\ref{fig:query-validation-tree} and~\ref{fig:recall-validation-tree}, parameters trained on \textit{Tree-$\alpha$} match its measurements closely, while those from \textit{ERBAC-$\alpha$} show mild deviations.

\begin{figure}[t]
  \centering
  \begin{minipage}{0.7\linewidth} 
    \centering
    \begin{minipage}{0.48\linewidth}
      \centering
      \includegraphics[width=\linewidth]{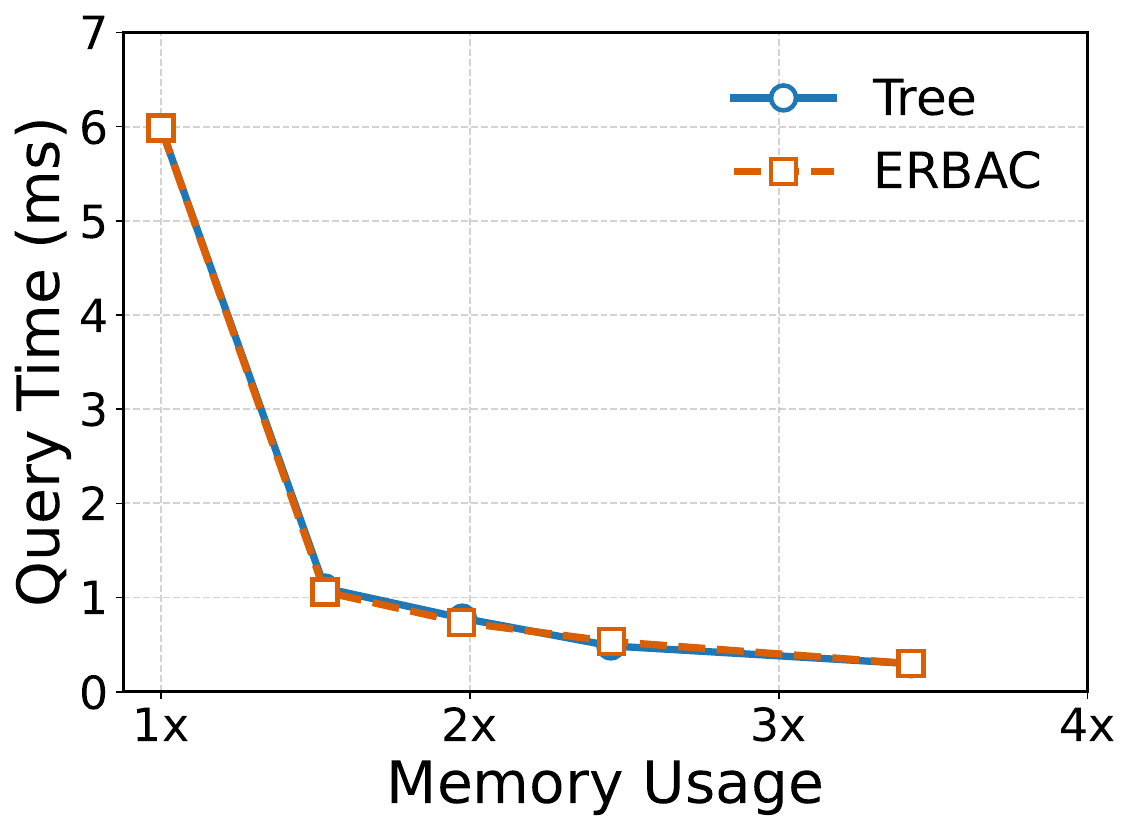}
      \caption{The effect of performance model on \sys.}
      \label{fig:performance-model-robustness}
    \end{minipage}
    \hfill 
    \begin{minipage}{0.48\linewidth}
      \centering
      \includegraphics[width=\linewidth]{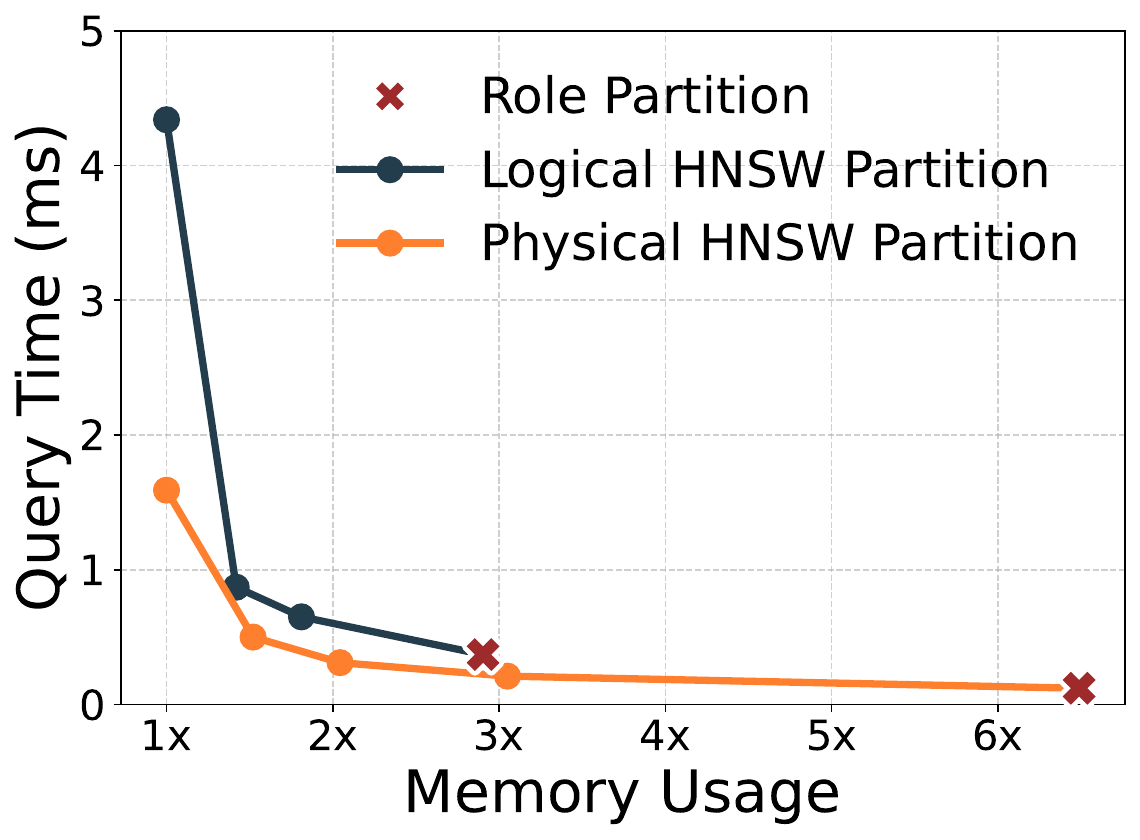}
      \caption{Comparison of logical vs physical HNSW partition (SIFT10M).}
      \label{fig:logical-vs-physical-partition}
    \end{minipage}
  \end{minipage}
\end{figure}

Second, we evaluate the effect of these parameter differences on \sys's partitioning algorithm.
Running \sys on \textit{Tree-$\alpha$} with both parameter sets yields nearly overlapping curves (Figure~\ref{fig:performance-model-robustness}), demonstrating that the partitioning algorithm is robust to slight parameter variations in the performance model. This is because even if the models are not perfectly accurate, they contain enough information to maintain the relative ranking among candidate partitions to guide optimization effectively.

\subsection{Physical vs Logical Partition}
\label{sec:eval-logical}
We compare two HNSW partitioning implementations.
In the \textit{physical} version, each partition stores both vectors and graph links.
The \textit{logical} version keeps only pointers to a shared vector table, reducing memory usage but introducing pointer indirection that increases query latency.

Logical partitioning still incurs nontrivial memory overhead from graph links and headers, which grows with the number of partitions.
Let $d$ be the vector dimension, $b_f$ the number of bytes per scalar (\eg $b_f{=}4$ if vectors and link IDs use float32/int32), and HNSW degree $M$ (each node keeps $\approx 3M$ links in total across levels, since it appears in $\approx$ 3 layers on average). The logical design uses
$b_f\!\left(|D|d+3M\sum_{\pi_j\in\Pi}|\pi_j|\right)$ bytes, compared to single-index baseline $b_f\!\left(|D|d+3M|D|\right)$. 
Thus the relative overhead is $\frac{|D|d+3M\sum_{j}|\pi_j|}{|D|d+3M|D|}$.

We modified \texttt{faiss} to support both implementations and evaluate on the Tree-$s$ workload (0.06 selectivity, SIFT10M dataset) with all optimizations disabled.
For this workload, Role Partition with physical partitions uses more than $6\times$ memory, while logical partitions reduce this to around $3\times$ the memory of a single index. 
As shown in Figure~\ref{fig:logical-vs-physical-partition}, physical partitioning consistently achieves lower latency under the same memory budget, providing a superior latency–memory trade-off. 
Logical partitioning is roughly 3$\times$ slower as each neighbor traversal accesses separate memory regions for graph and vector data, significantly increasing cache misses.


Logical partitioning can be beneficial in limited cases, such as with small datasets (hot caches), very high-dimensional vectors (where vector duplication signifantly increases memory cost), or extremely large numbers of roles (where one physical partition per role is costly).
Even then, \sys can optimize logical partitions to balance latency and memory effectively.

\subsection{Sensitivity Analysis}
\label{sec:sensitivity-analysis}
We analyze \sys's sensitivity to key parameters: dataset size, permission workload characteristics and top-$k$ values.

\begin{figure}[t]
  \centering
    \begin{minipage}{0.7\linewidth}  
    \centering
    \begin{subfigure}{0.48\linewidth}
    \centering
    \includegraphics[width=\linewidth]{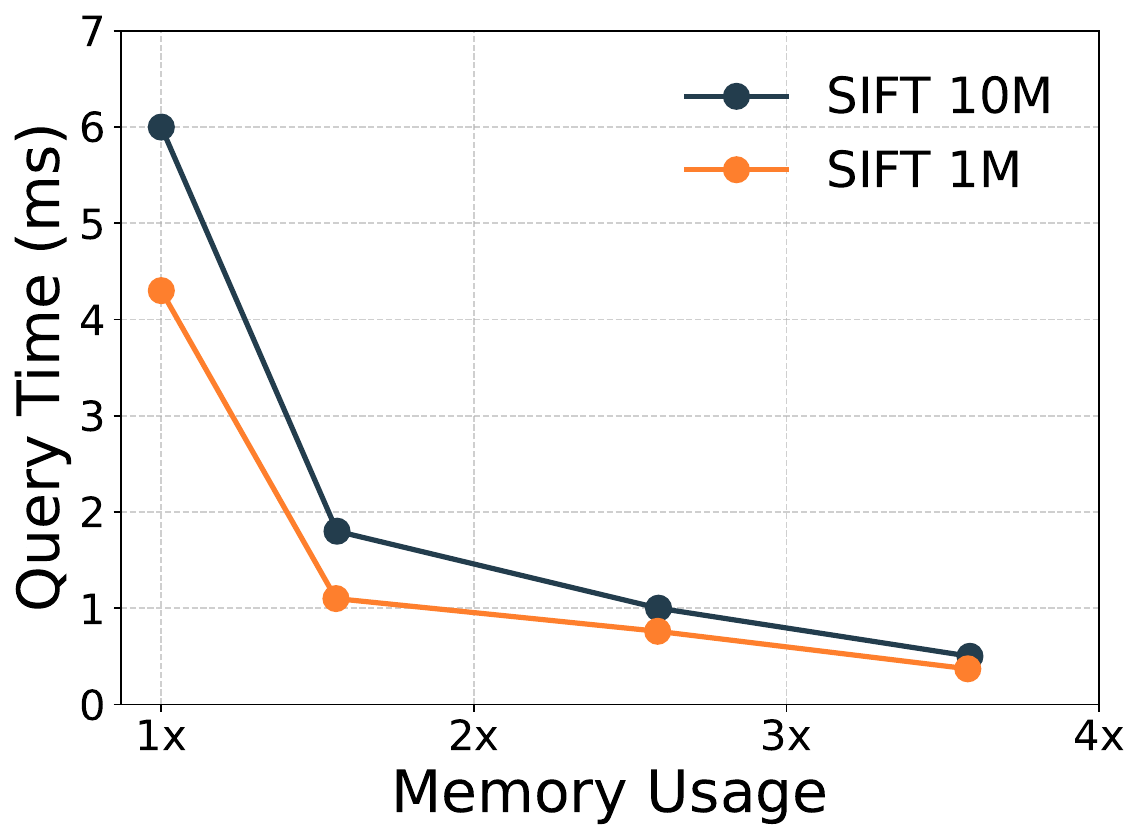}
    \caption{Impact of dataset size.}
    \Description{}
    \label{fig:data-scalability-comparison}
  \end{subfigure}
  \hfill
  \begin{subfigure}{0.48\linewidth}
    \centering
    \includegraphics[width=\linewidth]{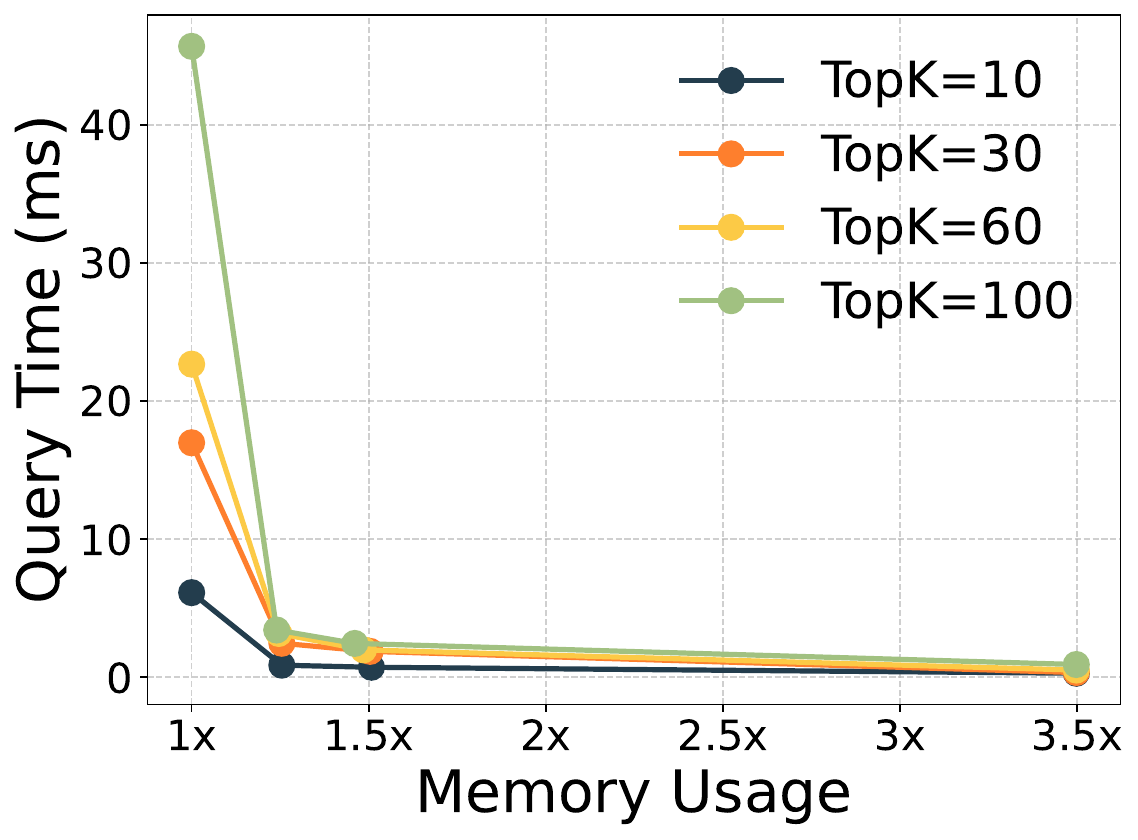}
    \caption{Impact of topk.}
    \Description{}
    \label{fig:topk-sensitivity-analysis}
  \end{subfigure}
  \vspace{-0.5em}
\caption{Sensitivity analysis on \sys performance.}
\Description{sensitivity_analysis}
\label{fig:sensitivity-analysis2}

\end{minipage}
\end{figure}


\minihead{Impact of Data Size}
We evaluate how dataset size (SIFT1M vs.\ SIFT10M) affects \sys's performance under different memory configurations using the Tree-$\alpha$ workload generator. As shown in Figure~\ref{fig:data-scalability-comparison}, increasing the dataset size degrades overall search performance, as expected, but the shape of the \sys trade-off curve remains largely unchanged.  
This suggests that data size primarily influences absolute latency, but the overall trade-off trend of \sys~remains~stable.


\minihead{Impact of topk}
We evaluated how different topk values (10, 30, 60, 100) affect \sys query performance across various memory configurations using Tree-$\alpha$ generator. Figure~\ref{fig:topk-sensitivity-analysis} shows that as k increases, RLS performance degrades significantly, while Role Partition maintains consistent performance. This sharp degradation occurs because RLS requires dramatically larger candidate sets (\efs > 3000 for k=100) to maintain recall as topk increases. 
\sys demonstrates a middle ground, with performance slowing as topk increases but at a much lower rate than RLS.

Importantly, the performance gap between RLS and \sys widens as topk increases, indicating \sys provides greater benefits for larger result sets. At higher memory constraints, \sys's performance degradation becomes less pronounced. For instance, with topk=100, \sys achieves approximately 13.5x faster query performance with only 1.24x memory compared to RLS and 90.4\% reduction in additional memory consumption compared to Role Partition (0.24$\times$ vs 2.5$\times$ additional), demonstrating excellent efficiency for high-topk retrieval scenarios.

\begin{figure}[t]
  \centering
    \begin{minipage}{0.7\linewidth}  
    \centering
  \begin{subfigure}{0.48\linewidth}
    \centering
    \includegraphics[width=\linewidth]{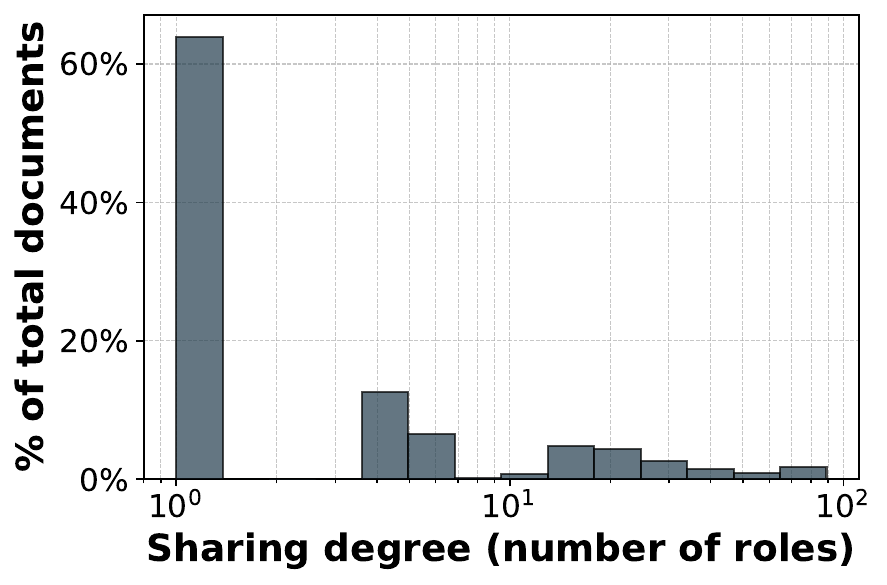}
    \caption{Sharing Degree Distribution Pattern (access selectivity 0.06).}
    \Description{}
    \label{fig:sharing-degree-treebased}
  \end{subfigure}
  \hfill
    \begin{subfigure}{0.48\linewidth}
    \centering
    \includegraphics[width=\linewidth]{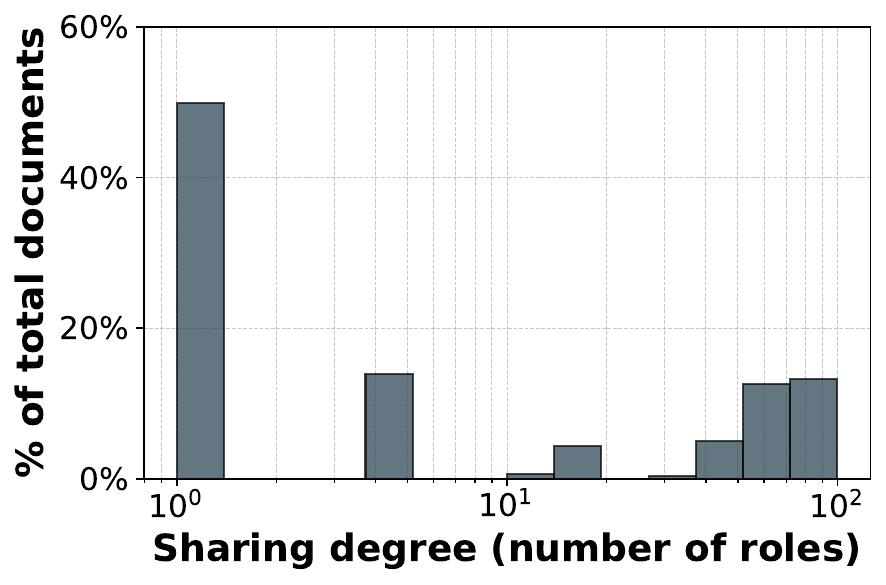}
    \caption{Sharing Degree Distribution Pattern (access selectivity 0.24).}
    \Description{}
    \label{fig:sharing-degree-treebased2}
    
  \end{subfigure}
\end{minipage}
\caption{Comparison of sharing degree distribution pattern in different access selectivity.}
\Description{sensitivity_analysis_selectivity}
\label{fig:similar-sharing-degree-pattern}
\end{figure}

\begin{figure}[t]
  \centering
    \begin{minipage}{0.7\linewidth}  
    \centering
    \begin{subfigure}{0.48\linewidth}
    \centering
    \includegraphics[width=\linewidth]
    {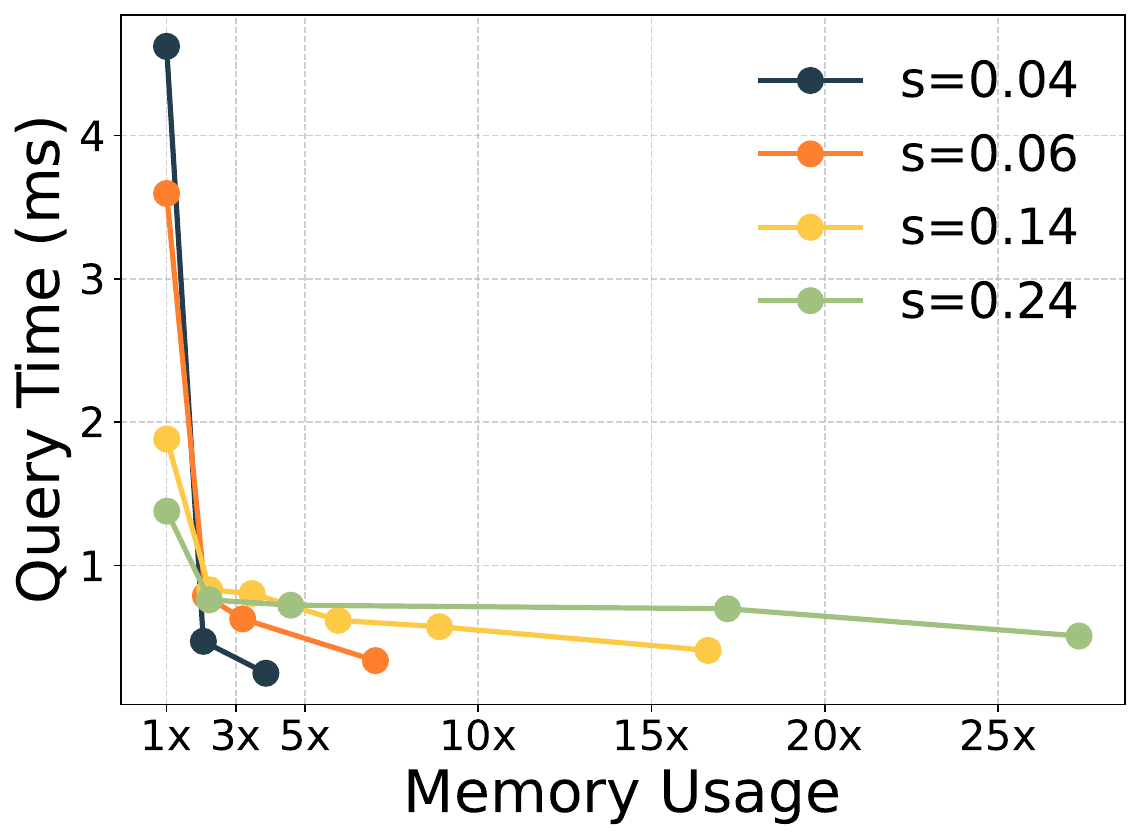}
    \caption{Effect of access selectivity}
    \Description{}
    \label{fig:sensitivity-analysis-selectivity}
  \end{subfigure}
\hfill
    \begin{subfigure}{0.48\linewidth}
    \centering
    \includegraphics[width=\linewidth]{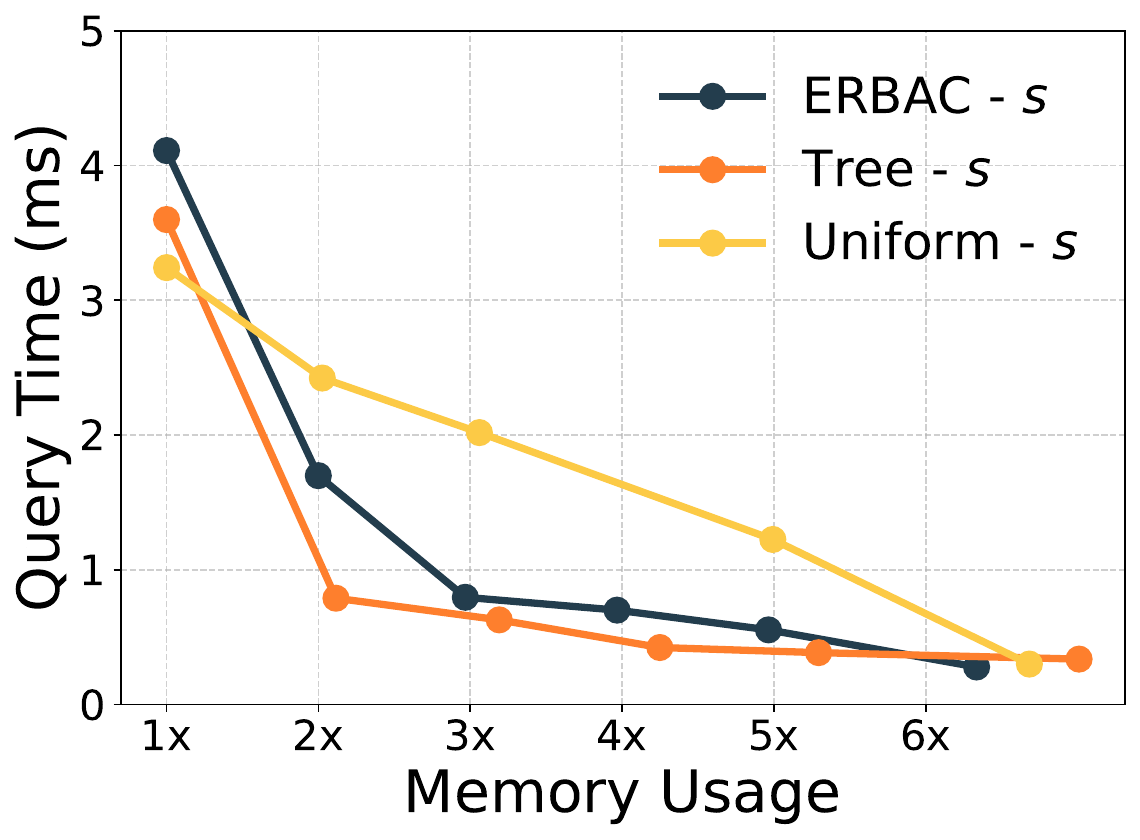}
    \caption{Effect of sharing degree pattern}
    \Description{}
    \label{fig:sensitivity-analysis-sharing-degree-pattern}
  \end{subfigure}
\end{minipage}
\caption{The effect of access selectivity and sharing degree distribution on \sys performance.}
\Description{sensitivity_analysis}
\label{fig:sensitivity-analysis3}
\end{figure}

\minihead{Impact of Access Selectivity}
We use the Tree-$s$ generator to examine sharing degree patterns at different access selectivity levels (0.04, 0.06, 0.14, 0.24). \autoref{fig:similar-sharing-degree-pattern} shows that while both selectivity levels exhibit similar patterns—sharing degree 1 has the highest peak with decreasing peaks for higher degrees—the magnitude differs: selectivity 0.06 (left) shows higher peaks than 0.24 (right), demonstrating consistent patterns with varying intensity.

\autoref{fig:sensitivity-analysis-selectivity} shows that at higher selectivity levels, RLS becomes more efficient, so the query latency at 1$\times$ memory approaches the performance of Role Partition (\ie $\Delta Latency$ is small). 
Additionally, Role Partition consumes more memory as selectivity increases ($\Delta Memory$). Therefore, \sys has the most optimization potential at low selectivity, where the gap between RLS and Role Partition ($\Delta Latency / \Delta Memory$) is largest.
For similar sharing degree distribution patterns, \sys's performance curves show similar trends at around 2$\times$ memory. 
We observe that the rate of improvement in query latency decreases the fastest at this point, whereas further increases in memory yield diminishing returns.

\begin{figure}[t]
  \centering
    \begin{minipage}{0.7\linewidth}  
    \centering
    \begin{subfigure}{0.49\linewidth}
    \centering
    \includegraphics[width=\linewidth]{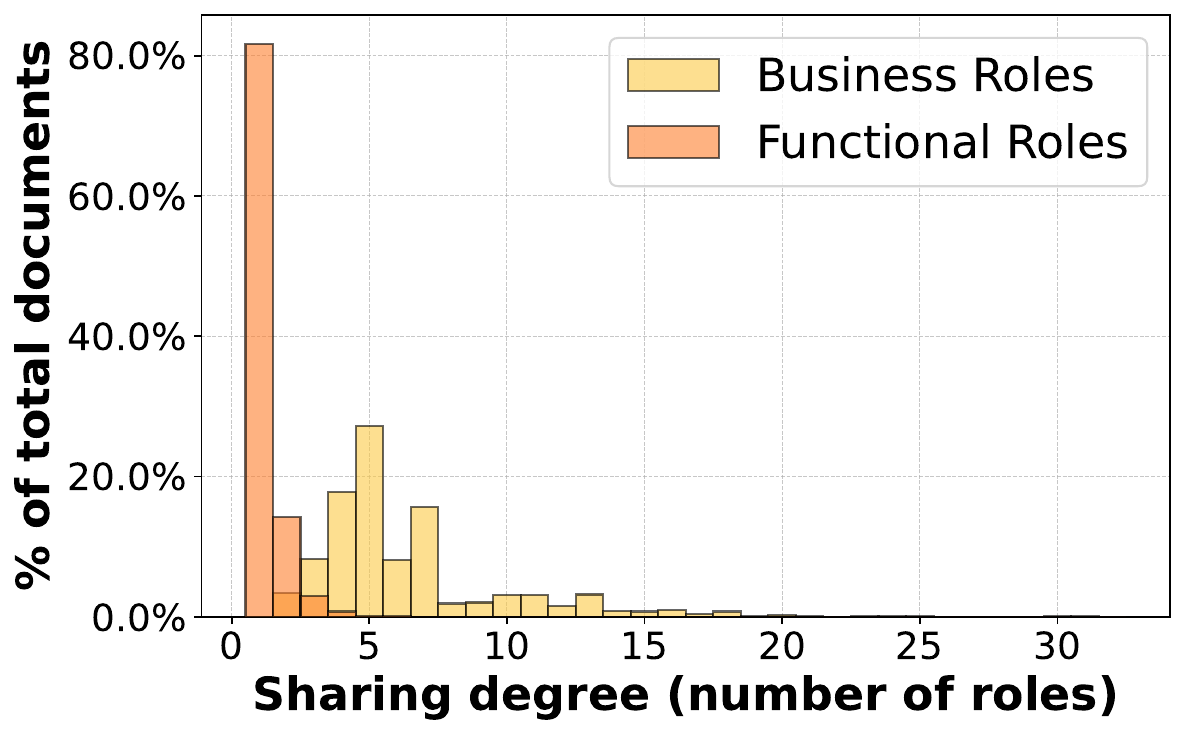}
    \caption{Sharing degree distribution pattern (ERBAC pattern).}
    \Description{}
    \label{fig:sharing-degree-erbac}
  \end{subfigure}
  \hfill
    \begin{subfigure}{0.46\linewidth}
    \centering
    \includegraphics[width=\linewidth]{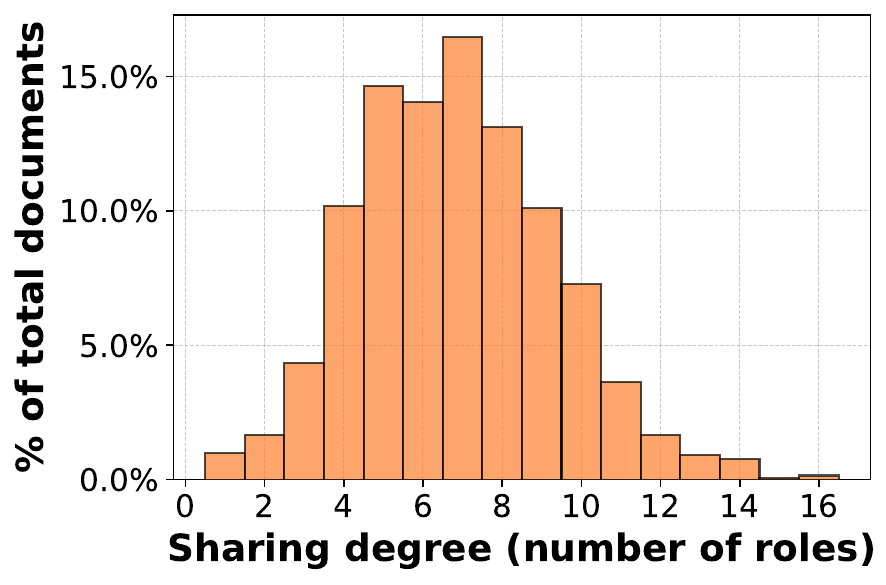}
    \caption{Sharing degree distribution pattern (Uniform pattern).}
    \Description{}
    \label{fig:sharing-degree-random}
  \end{subfigure}

\end{minipage}
\caption{Comparison of sharing degree distribution pattern in different permission workloads. }
\Description{sensitivity_analysis_sharing_degree}
\label{fig:different-sharing-degree-pattern}
\end{figure}

\minihead{Impact of Sharing Degree Pattern}
We evaluate \sys performance under workloads with different sharing degree patterns at similar selectivity (approximately 0.06 across cases).
Sharing degree distribution indicates document percentage accessible to varying role numbers (e.g., 50\% documents accessible to 1 role each, 20\% to 2 roles, etc.). 
Three distinct patterns are generated ensuring similar selectivity.
Tree pattern (\autoref{fig:sharing-degree-treebased}) peaks at small sharing degrees, indicating hierarchical structure where documents are primarily accessed by few roles.
ERBAC pattern (\autoref{fig:sharing-degree-erbac}) indicates two-layer hierarchical structure with functional and business roles. 
Uniform pattern (\autoref{fig:sharing-degree-random}), generated by the Uniform Generator, follows Poisson distribution with average sharing degree 7.
\autoref{fig:sensitivity-analysis-sharing-degree-pattern} demonstrates query latency and memory consumption across patterns are nearly identical at 1$\times$ memory (RLS), expected given workloads have identical average selectivity.

Overall, \sys enables different memory-query latency trade-offs, with performance ranking Tree $>$ ERBAC $>$ Uniform (evident from trade-off curve convexity). 
This demonstrates permission workload structure directly impacts \sys performance.
Comparing \autoref{fig:sharing-degree-treebased} and \autoref{fig:sharing-degree-erbac}, functional roles follow Tree pattern, but ERBAC performs worse due to the added complexity of business roles. Similarly, \autoref{fig:sharing-degree-erbac} vs. \autoref{fig:sharing-degree-random} demonstrates Uniform, lacking hierarchical structure, performs worse, particularly as its peak shifts from lower sharing degrees, making permission subset identification for partitioning more challenging.

\subsection{Adapting to Evolving Permission Workloads}
\label{sec:updateeval}
This section evaluates how \sys's partition design adapts to changes in permission workloads.

\begin{figure}[t]
  \centering
  \begin{minipage}{0.7\linewidth}  
\centering
  \begin{subfigure}
  {0.48\linewidth}
    \centering
    \includegraphics[width=\linewidth]{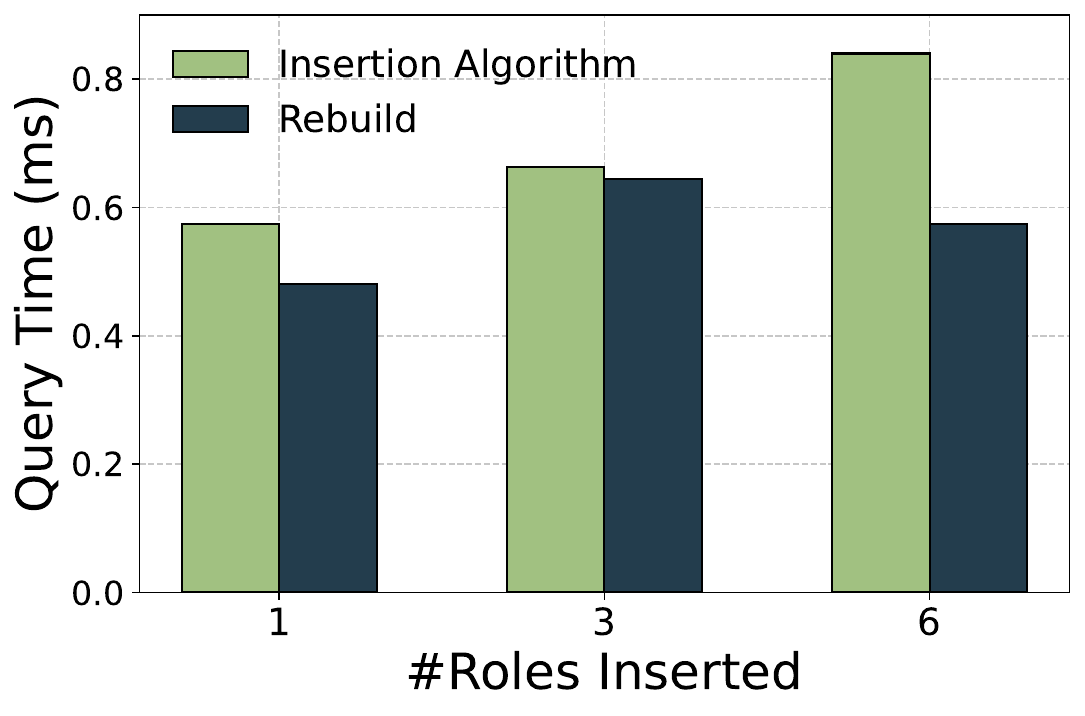}
    \caption{Insertion time benchmark.}
    \label{fig:insertion-time-benchmark}
  \end{subfigure}
  \hfill
  \begin{subfigure}{0.48\linewidth}
    \centering
    \includegraphics[width=\linewidth]{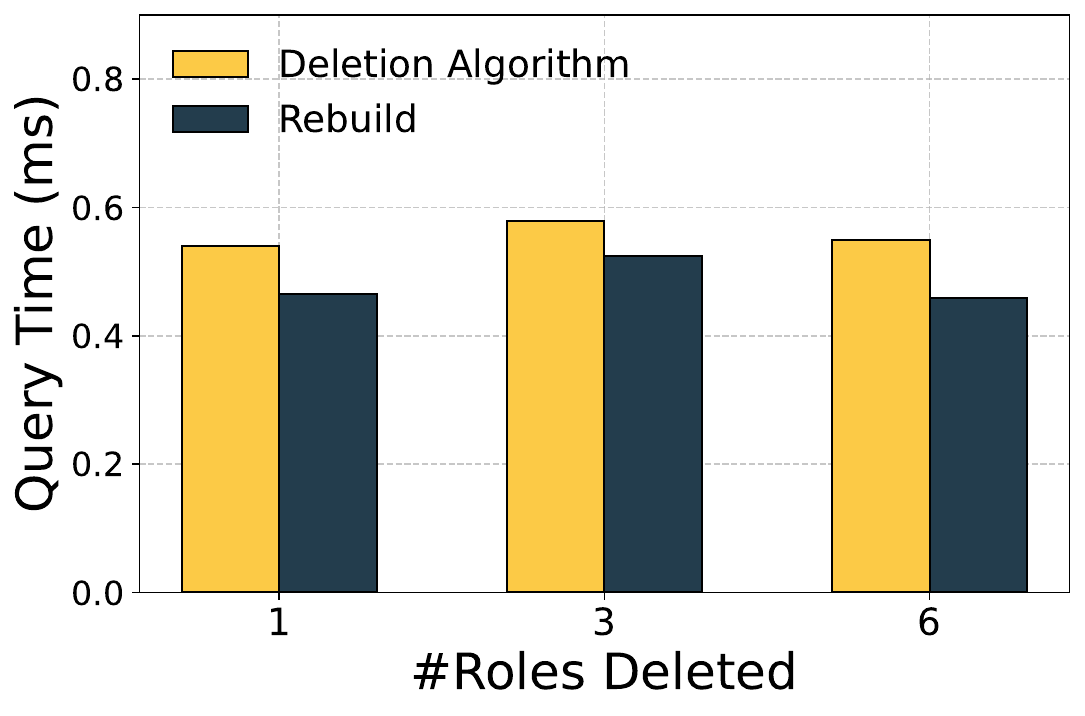}
    \caption{Deletion time benchmark.}
    \label{fig:deletion-time-benchmark}
  \end{subfigure}
\end{minipage}
\caption{Query latency comparison: incremental partition maintenance vs. full rebuild under permission workload changes. }
  \label{fig:Updating-benchmark}
\end{figure}

We first evaluate how efficiently \sys maintains its partitioning strategy during permission workload changes, specifically role insertions and deletions. We use Tree-$\alpha$ generator with \sys at 1.5$\times$ memory. Insertions create new roles from existing document subsets, adding users equal to 1\% of the original user base. Deletions remove roles and their exclusive users. We use one-to-one user-role mapping and compare incremental updates to full rebuilds across 1, 3, or~6 operations.

\autoref{fig:insertion-time-benchmark} shows that for few insertions (<5), incremental updates perform similarly to rebuilds. However, as insertions increase, incremental partitioning shows higher query latency, suggesting \sys benefits from rebuilding when workloads change significantly.
\autoref{fig:deletion-time-benchmark} shows incremental deletion consistently maintains query latency comparable to complete rebuilds. Role modifications naturally change total memory footprint for all methods. Based on our observation, while absolute memory consumption changes, the relative memory overhead ratios among RLS, \sys, and Role Partition remain consistent with those reported in Section~\ref{sec:main-results}.




\section{Related Work}
\label{sec:related}

\minihead{Partitioning in Databases}
Traditional horizontal partitioning methods distribute data by key attributes to enable partition pruning, parallel processing, and load balancing~\cite{ceri1982horizontal, cpalka2014newhorizontal, agrawal2004integratinghorizontal, bellatreche2000algorithmshorizontal,dimovski2010horizontal,zilio1998physical}. For example, QD-tree and followup works~\cite{yang2020qd, mto,pando,rong2024dynamic} optimizes data layouts by partitioning based on query patterns to minimize the number blocks accessed.
However, these approaches target traditional query patterns--exact-match or range predicates--rather than similarity search with permission filtering, where vector distance calculations combined with permission checks create unique performance characteristics. HQI~\cite{hqi} extends QD-trees to similarity search queries.
It leverages query-workload predicates to build balanced QD-trees and construct separate IVF indexes per partition, batching queries by shared attribute predicates. 
HQI can be considered as a special case where memory overhead is 1$\times$, while \sys optimizes partitioning for any memory constraint ($\geq 1\times$) through overlapping partitions. As shown in Figure~\ref{fig:query-time-vs-storage}, HQI’s performance is highly unpredictable: it performs well in certain case (e.g., Tree-$\alpha$) but degrades sharply in others (e.g., Uniform or ERBAC). This inconsistency arises because HQI relies purely on heuristic partitioning without any performance-guided model. To the best of our knowledge, \sys is the first system to systematically explore strategic vector replication across partitions achieving efficient memory-performance tradeoffs in vector databases, formulating this as constrained optimization with explicit memory constraints. 

\minihead{Hybrid Search Indexes} 
Hybrid search indexes integrate structured filtering into vector similarity search for efficient constraint enforcement like access control~\cite{patel2024acorn,gollapudi2023filtered,irange,wu2022hqann,wang2024efficient}. ACORN~\cite{patel2024acorn} provides performant, predicate-agnostic hybrid search handling high-cardinality and unbounded predicate sets. It improves HNSW by integrating filtering capabilities directly into the graph structure while remaining compatible with existing HNSW implementations. HQANN~\cite{wu2022hqann} provides hybrid query processing framework embedded into proximity graph-based ANN algorithms. It manages hybrid queries through unified processing, enhancing performance and adaptability.
Filtered-DiskANN~\cite{gollapudi2023filtered} constructs graph-structured indexes leveraging vector geometry and label metadata (e.g., date, price). This approach incorporates geometric relationships and associated labels for effective navigational graph structure.
Our partitioning design is orthogonal to index type. As described in \S~\ref{sec:exp-index}, it complements hybrid search by addressing partitioning and access policy management challenges, enabling integration of hybrid search techniques like ACORN with~our~method.

\begin{table}[t]
\centering
\footnotesize
\caption{Overview of related work in hybrid vector search.} 
\label{tab:related}
\begin{tabular}{lclc}
\toprule
& \textbf{Method type} & \textbf{Index agnostic?} & \textbf{Memory constraint?}  \\
\midrule
\textbf{ACORN}~\cite{patel2024acorn} & index & $\times$ (HNSW) & $\times$\\
\textbf{Filtered-DiskANN}~\cite{gollapudi2023filtered} & index & $\times$ (Vamana) & $\times$\\
\textbf{Curator}~\cite{jin2024curator} & index & $\times$ (IVF) & $\times$ \\
\textbf{HQI}~\cite{hqi} & partition & $\checkmark$ & $\times$ \\
\textbf{\sys} (ours) & partition & $\checkmark$ & $\checkmark$ \\
\bottomrule
\end{tabular}
\end{table}

\minihead{Multi-Tenant Vector Databases} 
Curator~\cite{jin2024curator} represents the most relevant prior work in vector database access control for multi-tenant scenarios.
Multi-tenant database implementations generally follow three patterns~\cite{hui2009supporting}, with different trade-offs in data isolation and resource usage: (1) isolated db instances per tenant, (2) per-tenant tables within a shared db instance (\ie the specialized index approach), and (3) shared tables storing all tenant data with tenant identification columns (\ie the post-filtering approach). 
At a high level, Curator introduces a multi-tenant index to improve approach (3), while \sys helps users navigate the trade-off between approaches (2) and (3) for a given index.

Specifically, Curator introduces a hierarchical k-means clustering tree tailored to each tenant’s vector distribution, and embeds permission filters using Bloom filters within a shared clustering tree, enabling efficient search by skipping inaccessible vector clusters with minimal memory overhead.
Curator differs from \sys in two key aspects: (1) Curator uses direct user-permission mappings (ACLs), while \sys leverages RBAC hierarchies for scalable partitioning; (2) Curator's ACL logic is tightly coupled with its custom k-means tree design, limiting reuse of standard ANN indexes, while \sys separates partition design from index implementation, enabling seamless integration with HNSW, IVF-Flat, and hybrid techniques.
Similar to ACORN~\cite{patel2024acorn}, Curator could integrate \sys's dynamic partitioning within its k-means clustering tree to gain further performance improvement.

\section{Conclusion}
\label{sec:conclusion}
We presented \sys, a dynamic partitioning framework for access control in vector databases. \sys is the first to systematically explore strategic vector replication across partitions to achieve efficient memory–performance tradeoffs, formulating partitioning as a constrained optimization problem with explicit memory limits.
Leveraging RBAC structure, \sys partitions the vector space to minimize query latency, guided by analytical models predicting search performance and recall. Experiments show up to 13.5$\times$ faster queries than row-level security with only 1.24$\times$ more memory, and 90.4\% less memory than role partitioning at comparable speed. 
\sys offers a practical, scalable solution for enforcing access control in vector databases.

\begin{acks}
This work was supported in part by the National Science Foundation under grant IIS-2335881.
\end{acks}



\bibliographystyle{ACM-Reference-Format}
{ 
\raggedright 
\bibliography{confs_long, main}
}

\received{July 2025}
\received[revised]{October 2025}
\received[accepted]{November 2025}

\appendix

\section{Optimization Problem Formulation}
\label{sec:optimization}
Given the analytical models for query performance and recall, we can formulate Problem~\ref{def:prob} as a constraint optimization problem.

\noindent\textbf{Given Constants:}
\squishitemize
\item $U, R, D$: Sets of users, roles, and documents.
\item $auth(u_i):$ User-to-document access mapping from RBAC (Def~\ref{def:rbac}).
\item $topk$: A representative \topk value\footnote{Although $k$ is a required input for the recall model, in the evaluation (\S~\ref{sec:robusttopk}), we show that the partitioning design is not sensitive to the specific $topk$ used during the optimization process.}.
\item $n_p$: Number of partitions in configuration $\Pi$.
\item $a,b$: Fitted parameters for the query performance model (Eq~\ref{eq:usercost}).
\item $R(\efs, s, topk)$: Recall model with fitted parameters (Eq~\ref{eq:recalldef}).
\item $\alpha$: Memory overhead constraint ($\geq 1$).
\item $\epsilon$: Minimum recall threshold ($< 1$).
\squishend

\noindent\textbf{Variables:}
\squishitemize
\item $p_{j,k} \in \{0,1\}^{|D| \times n_p}$: Binary decision variable indicating whether document $d_j$ is assigned to partition $\pi_k$. 
\item \(x_{i,k} \in \{0,1\}^{|U| \times n_p}\): Binary decision variable indicating whether user $u_i$ should access partition $\pi_k$ for their queries. 
\item $\efs$: HNSW search depth parameter
\squishend

The objective is to minimize the average query cost across all users, based on the user-level performance model (Eq~\ref{eq:usercost}). The constraints ensure that: (1) user access indicators \(x_{i,k}\) accurately reflect document assignments via \(\text{acc}(u_i)\), using a helper variable \(\delta_{i,j,k}\) to link \(p_{j,k}\) and \(x_{i,k}\) (Eq~\ref{eq:useraccess}, ~\ref{eq:deltaaccess},~\ref{eq:deltax}); during optimization, \(x_{i,k}\) naturally approaches \(\text{P}^{*}\) defined in Eq~\ref{eq:trackermin}, (2) no partition is empty (Eq~\ref{eq:doccheck}), (3) total memory overhead doesn't exceed $\alpha$ (Eq~\ref{eq:storage}), (4) recall meets the minimum threshold $\epsilon$ using the model from Eq~\ref{eq:recalldef} (Eq~\ref{eq:recall}), and (5) average access selectivity is properly computed based on document assignments and access patterns (Eq~\ref{eq:sel}).

\begin{mini!}|l|[3]<l>
  {p_{j,k},x_{i,k}, \efs}{\frac{1}{|U|}\sum_{i=1}^{|U|}\sum_{k=1}^{n_p} x_{i,k}\cdot \log\left(\sum_{j=1}^{|D|} p_{j,k} \right) (a \cdot \efs + b)}
  {\label{eq:opt}}{}
  \addConstraint{\sum_{k=1}^{n_p} x_{i,k} \geq 1, \quad \forall i \in \{1, 2, \ldots, |U|\}}{\label{eq:useraccess}\quad\text{(access mapping)}}
  \addConstraint{\sum_{k=1}^{|D|} \delta_{i,j,k} \geq 1, \quad \forall i \in \{1, 2, \ldots, |U|\},\forall j \in auth(u_{i}) }{\label{eq:deltaaccess}}
  \addConstraint{x_{i,k} \geq \delta_{i,j,k}, \delta_{i,j,k} \leq p_{j,k} \quad \forall i, j, k}{\label{eq:deltax}}
  \addConstraint{\sum_{j=1}^{|D|} p_{j,k} \geq 1, \quad \forall k \in \{1,\ldots,n_p\}}{\label{eq:doccheck}\text{(non-empty partitions)}}
  \addConstraint{\sum_{k=1}^{n_p} \sum_{j=1}^{|D|} p_{j,k} \leq \alpha |D|}{\label{eq:storage}\quad\text{(memory)}}
  \addConstraint{R(\efs, \overline{s}, topk) \geq \epsilon}{\label{eq:recall}\quad\text{(recall)}}
  \addConstraint{\overline{s} = \frac{1}{|U|}\sum_{i=1}^{|U|}\frac{1}{\sum_{k=1}^{n_p}x_{i,k}} \sum_{k=1}^{n_p}x_{i,k}\cdot \frac{\sum_{j \in auth(u_i)}p_{j,k}}{\sum_{j}p_{j,k}  }}{\label{eq:sel}\text{(access selectivity})}
\end{mini!}

The optimization process involves three steps. 
First we compute the user's access selectivity $s(u_i)$ and the average access selectivity $\overline{s}$ from document assignments $p_{j,k}$. 
Second, we determine the minimum $\efs$ needed to achieve recall threshold $\epsilon$; we apply the same $\efs$ for all partitions. 
Finally, we optimize partition assignments $p_{j,k}$ to minimize average query time.

This optimization problem belongs to the class of Mixed-Integer Nonlinear Programming (MINLP), which is NP-hard due to its combinatorial nature and nonlinear constraints. 
In addition, the the problem involves $O((|D|+|R|)n_p)$ binary variables, making it intractable for large-scale datasets using off-the-shelf solvers. 
Our greedy dynamic partitioning algorithm effectively solves this optimization problem.
While \(n_p\) (number of partitions) is treated as a constant in the optimization problem, the greedy algorithm treats \(n_p\) as a variable to enhance flexibility.

\section{Supporting Hybrid Filtering Modes}

In this section, we discuss how to extend \sys to support pre-filtering for high-selectivity scenarios.
We discuss two possible approaches.

\paragraph{Online switching without repartitioning.}
AnonySys can directly reuse the partitions produced offline and, at \emph{query time}, choose the filtering mode based on the estimated effective selectivity $\overline{s_u}$ of the target role/user.
Concretely, low-selectivity queries use HNSW with post-filtering, while highly selective queries first apply predicate pre-filtering and then perform brute-force distance evaluation on the filtered subset.
This online switch requires no repartitioning and is orthogonal to the offline partition design.

\paragraph{Partition-aware switching via a piecewise model.}
If we want the \emph{partitioning} itself to be aware of the switch, we can equip the optimizer with a piecewise performance model that captures the crossover between post- and pre-filtering:

\[
C(u,\pi) \;=\;
\begin{cases}
O\!\Big(\tfrac{k}{\overline{s_u}(\pi)} \cdot \log|\pi|\Big),
& \text{if } \overline{s_u}(\pi) \le s_{\text{th}}(\pi),\\[8pt]
O\!\Big(c_{\text{pred}}\,|\pi| \;+\; c_{\text{bf}}\,\overline{s_u}(\pi)\,|\pi|\Big),
& \text{if } \overline{s_u}(\pi) > s_{\text{th}}(\pi),
\end{cases}
\]

where $c_{\text{pred}}$ is the per-vector predicate evaluation cost for pre-filtering, and $c_{\text{bf}}$ is the per-distance cost for brute-force evaluation on the filtered subset of size $\overline{s_u}(\pi)\,|\pi|$.
The threshold $s_{\text{th}}(\pi)$ marks the selectivity at which the two modes have equal expected cost:

\[
\tfrac{k}{s}\log|\pi| \;=\; c_{\text{pred}}\,|\pi| + c_{\text{bf}}\,s\,|\pi|
\quad\text{at } s = s_{\text{th}}(\pi).
\]

This hybrid formulation enables the optimizer to reason jointly about memory, selectivity, and filtering strategy, while keeping the pre-filter branch faithful to a brute-force backend after predicate pruning.





\section{Additional Evaluation}
\label{sec:robusttopk}
To evaluate the robustness of \sys's partition schemes to variations in \topk values during query time, we compare two partition schemes: one optimized for \topk=10 and another for \topk=100, both using 1.5$\times$ memory and 0.9 recall constraints using Tree-$\alpha$ generator.

As shown in Table~\ref{tab:topk-partition-comparison}, the \topk=10 scheme created 14 partitions (1939MB), while \topk=100 produced 15 partitions (1916MB). 
When evaluated on the same 1000 test queries with \topk=100, both achieved similar performance: the \topk=10 scheme yielded 2.67ms query time with 0.990 recall, compared to 2.87ms and 0.991 recall for the \topk=100 scheme.
This minimal difference (less 7\% in query time) demonstrates that \sys partition schemes are robust to \topk variations, allowing \sys to effectively handle queries requesting different numbers of results without the need to rebuild partitions for new \topk values.

\begin{table}[t]
\small
\centering
\caption{Comparison of partition schemes generated with different \topk values.}
\label{tab:topk-partition-comparison}
\begin{tabular}{ccccc}
\toprule
\textbf{Generation} & \textbf{Number of} & \textbf{Memory} & \textbf{Avg Query} & \textbf{Avg} \\
\textbf{\topk} & \textbf{Partitions} & \textbf{Used (MB)} & \textbf{Time (ms)} & \textbf{Recall} \\
\midrule
10 & 14 & 1939 & 2.67 & 0.990 \\
100 & 15 & 1916 & 2.87 & 0.991 \\
\bottomrule
\end{tabular}
\end{table}

\end{document}